\documentstyle[aps,epsf,eqsecnum,tighten,preprint,floats]{revtex}
\begin{document}
\title{Intermediate temperature dynamics of one-dimensional Heisenberg
antiferromagnets}
\author{Chiranjeeb Buragohain and Subir Sachdev}
\address{Department of Physics, Yale University\\
P.O. Box 208120,
New Haven, CT 06520-8120, USA}
\date{Nov 5, 1998}
\preprint{cond-mat/9811083}
\maketitle
\begin{abstract}
We present a general theory for the intermediate temperature ($T$)
properties of Heisenberg antiferromagnets of spin-$S$ ions on
$p$-leg ladders, valid for $2Sp$ even or odd. Following an earlier
proposal for $2Sp$ even (Damle and Sachdev, Phys. Rev. B {\bf 57}, 8307), we
argue that an integrable, classical, continuum model of a fixed-length, 3-vector
applies over an intermediate temperature range; this range
becomes very wide for
moderate and large values of $2Sp$.
The coupling constants of the effective model are
known exactly in terms of the energy gap above the ground
state $\Delta$ (for $2 S p$ even), or a crossover scale $T_0$ (for $2 S p$
odd). Analytic and numeric results for dynamic and transport
properties are obtained, including some exact results for the
spin-wave damping. Numerous quantitative predictions for neutron scattering
and NMR experiments are made.
A general discussion on the nature of $T>0$
transport in integrable systems is also presented:
an exact solution of a toy model proves
that diffusion can exist in integrable systems, provided proper care
is taken in approaching the thermodynamic limit.
\end{abstract}
\pacs{PACS numbers:}
%\debugon
\newpage
\section{Introduction}
\label{intro}
One-dimensional Heisenberg antiferromagnets are strongly-interacting
quantum many body systems for which a detailed quantitative
confrontation between theory and experiment has been possible.
A rather precise, and parameter-free understanding of their
low temperature dynamic properties has emerged in a number of
recent NMR experiments performed by Takigawa and
collaborators~\cite{taki1,taki2,taki3,kedar}. These systems
can therefore serve as useful springboards towards deciphering the
behavior of interacting systems of greater complexity.

The past theoretical work on the dynamic properties of these
quantum antiferromagnets has focused mainly only the universal behavior
in the asymptotic
low temperature ($T$) regime $T \rightarrow 0$~\cite{schulz,ssnmr,star1,star2,kedar}.
In the present paper we will extend the theory to a separate range
of intermediate temperatures.
We shall argue that under suitable
conditions, to be described precisely below, this intermediate
temperature range can be quite wide, and is described by a
continuum dynamical model quite different from that required
for $T\rightarrow 0$. This intermediate temperature dynamics was
discussed briefly for a limited class of antiferromagnets
in the last section of Ref~\onlinecite{kedar}.

Our work will also connect with earlier investigations of the
dynamics of classical lattice antiferromagnets (Refs~\cite{blume,reiter} and
references therein). In a sense, our paper provides a bridge
between the modern quantum dynamics
and the classical studies of the 70's. There is an overlapping
window of validity for our theory and the classical investigation of Reiter and
Sj\"{o}lander~\cite{reiter}, and here our results are generally
consistent with theirs, although there are some details that disagree. We will review
this earlier work, in the context of our results, in
Section~\ref{expts}.

A large fraction of the
experimental examples of one-dimensional Heisenberg antiferromagnets
consist of $p$ parallel, coupled chains of spin $S$ ions
(for $p=1$ these are ordinary
spin chains, while for $p>1$ these are commonly referred to as
$p$-leg ladders).
For all $T < T_{\rm max}^{(1)}$, where $T_{\rm max}^{(1)}$ will be
defined shortly, these antiferromagnets are described by a
universal quantum field theory: the one-dimensional
${\rm O} (3)$ non-linear sigma model. This field theory has the
quantum partition function (in units with $\hbar = k_B =1$, which we
use throughout)
\begin{eqnarray}
&& {\cal Z}_Q = \int {\cal D} {\bf n} (x, \tau) \delta( {\bf n}^2 - 1)
\exp \left(- \int d x \int_0^{1/T} d \tau \, {\cal L} \right)
\nonumber \\
&& {\cal L} = \frac{1}{2 c g} \left[
\left(\frac{\partial {\bf n}}{\partial \tau} - i {\bf H} \times
{\bf n} \right)^2 + c^2 \left( \frac{\partial {\bf n}}{\partial x}\right)^2 \right]
+ \frac{i \theta}{4 \pi} {\bf n} \cdot \left( \frac{\partial {\bf
n}}{\partial x}
\times \frac{\partial {\bf n}}{\partial \tau} \right).
\label{action}
\end{eqnarray}
Here ${\bf n} (x, \tau)$ is a three-component unit vector
representing the orientation of the antiferromagnetic order
parameter at spatial position $x$ and imaginary time $\tau$,
$c$ is a spin-wave velocity, and ${\bf H}$ is a uniform external
magnetic field---we will be interested only in the linear response
to ${\bf H}$. There are two dimensionless
coupling constants in ${\cal L}$, $\theta$ and $g$.
The first, $\theta$, is the coefficient of a topological term,
and has the value $\theta = \pi$ for $2Sp$ an
odd integer, and the spectrum
of excitations above the ground state is then gapless.
For $2 S p$ an even integer, $\theta=0$, and then there is gap to
all excitations. The coupling $g$ plays a role
in determining the energy scale at which certain crossovers (to be
discussed below) take place, but does not modify the physics
otherwise. A straightforward semiclassical (large $S$)
derivation shows that
\begin{eqnarray}
g &\approx& \frac{2}{Sp} \left[ 1 + \left(1 - \frac{1}{p} \right)
\frac{J_{\perp}}{J} \right]^{1/2} \nonumber \\
c &\approx& 2 J S a \left[ 1 + \left(1 - \frac{1}{p} \right)
\frac{J_{\perp}}{J} \right]^{1/2},
\label{gest}
\end{eqnarray}
where $J$ is the exchange constant along the legs of the ladder,
$J_{\perp}$ is exchange on the rungs, and $a$ is the lattice
spacing along the legs; we have assumed
here a model with only nearest neighbor exchange, but the
estimates $g \sim 1/Sp$ and $c \sim JSa$ hold far more generally.
An important observation is that $g$
becomes small for either large $S$ or $p$. We will be
especially interested in the small $g$ case in this paper.

Let us now discuss the value of $T_{\rm max}^{(1)}$ below
which (\ref{action}) holds. The basic argument follows that made
by Elstner {\em et al\/}~\cite{nobert} in $d=2$. At a temperature
$T$, the characteristic excited spin-wave has wavelength $c/T$,
and the continuum quantum theory will apply as long as this
wavelength is longer than the lattice spacing, $a$, of the
underlying antiferromagnet. For $p$-leg ladders, description by a
one-dimensional quantum model requires that the wavelength be
larger than the width of the ladder, $pa$~\cite{olaf}.
Using the value of $c$ in (\ref{gest}),  our estimate
for $T_{\rm max}^{(1)}$ is then
\begin{equation}
T_{\rm max}^{(1)} \sim \frac{ 2 J S}{p} \left[ 1 + \left(1 - \frac{1}{p} \right)
\frac{J_{\perp}}{J} \right]^{1/2}.
\label{t1est}
\end{equation}
To reiterate, the quantum theory (\ref{action}) applies to the
lattice antiferromagnet at all $T$ below that in (\ref{t1est}).

\begin{figure}
\epsfxsize=4in
\centerline{\epsffile{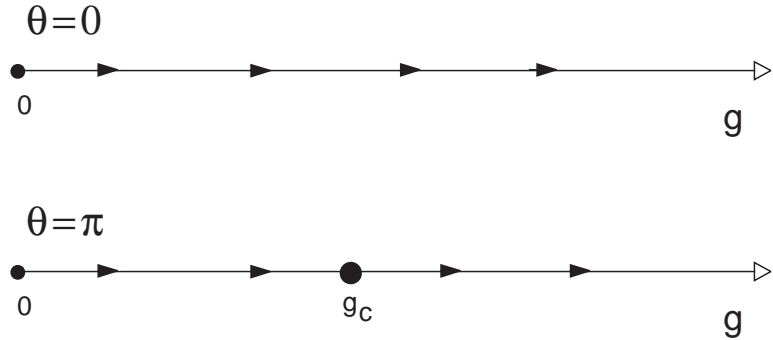}}
\caption{Renormalization group flows for the dimensionless coupling $g$
in (\protect\ref{action}).
For $\theta = 0$, $g$ has a runaway flow to $g=\infty$ and the
ground state is a quantum paramagnet with a gap $\Delta$.
For $\theta = \pi$, there is a fixed point at
$g = g_c$, of order unity,
and near it the flow is $dg/d\ell \propto (g-g_c)^2$. This fixed
point is described by the $k=1$, $SU(2)$ Wess-Zumino-Witten model.
The crossover between the $g=0$ and $g=g_c$ fixed points takes
place at an energy scale of order $T_0$. The region $g>g_c$
usually corresponds to a gapped state with spin-Peierls order,
and is not considered in this paper.
}
\label{fig1}
\end{figure}

Let us now review the well-known, $T=0$,
renormalization group
properties of (\ref{action})~\cite{affhal}. The topological angle $\theta$ remains
fixed at $\theta=0,\pi$, while the flows of the coupling
$g$ are sketched in Fig~\ref{fig1}.
For both cases $\theta=0,\pi$ there is a fixed point $g=0$
which is unstable at low energies. Indeed the beta function
describing the flow away from $g=0$ is independent of $\theta$ to
all orders in $g$. However, non-perturbative, topological effects
do distinguish the two values of $\theta$. For $\theta =0$, the
flow is believed to continue all the way to $g = \infty$,
corresponding to a quantum paramagnetic ground state with an
energy gap $\Delta$. In contrast, for $\theta = \pi$, the flow is
into a strong coupling, infrared stable fixed point at
$g=g_c$. There is a scale-invariant and gapless theory which
describes this fixed point---the $k=1$, $SU(2)$,
Wess-Zumino-Witten model. For $g>g_c$, there is a runaway flow to
$g=\infty$, usually associated with the appearance of
spin-Peierls order; this last regime will not be discussed in this
paper.

Our primary interest here shall be in the region in the vicinity
of the unstable $g=0$ fixed point. For both $\theta = 0, \pi$,
there is a characteristic energy scale, usually denoted
$\Lambda_{\overline{MS}}$, which determines the location of the
crossover from the vicinity of the $g=0$ fixed point to the
strong-coupling behavior. For energies or temperatures
smaller than $\Lambda_{\overline{MS}}$, the strong coupling
behavior should apply, and as we have just discussed, this is
quite different for $\theta = 0$ and $\theta = \pi$. However, the
physics at energies or temperatures larger than $\Lambda_{\overline{MS}}$
is controlled by the flow in the vicinity of the $g=0$ fixed
point, and this is common to both $\theta =0 $ and $\theta = \pi$.
The energy scale $\Lambda_{\overline{MS}}$ can be estimated from
the structure of the perturbative $\beta$-function. For small $g$
\begin{equation}
\Lambda_{\overline{MS}} \sim J \exp\left( - \frac{2 \pi}{g}
\right);
\label{MS}
\end{equation}
we have neglected here a prefactor of a power of $g$ coming from
higher-loop corrections. So from (\ref{gest}), for either $S$ or $p$
moderately large, the scale $\Lambda_{\overline{MS}}$ becomes
{\em exponentially small.}

We are now ready to discuss the static thermodynamic properties of
the quantum field theory (\ref{action}) as a function of $T$.
We will characterize the system by $T$ dependence of
two important observables: $\chi_u (T)$ and $\xi (T)$.
The first is the uniform susceptibility, $\chi_u$,
which is the linear response to the field ${\bf H} =
(0,0,H)$: $\chi_u = (T/ L) (d^2 \ln Z_Q / d H^2 ) |_{H=0}$,
where $L$ is the (infinite) length of the spatial direction; this
is the susceptibility {\em per rung} of the ladder.
The second is the correlation length, $\xi (T)$, which determines
the exponential decay of the equal-time two-point ${\bf n}$ field
correlator as a function of $x$.

We will consider temperatures above and below $\Lambda_{\overline{MS}}$
in turn.

\noindent
(A) \underline{$T < \Lambda_{\overline{MS}}$}\\
For $T < \Lambda_{\overline{MS}}$, as just noted, we must
distinguish $\theta = 0$ and $\theta = \pi$.

\noindent
(A.I)\underline{$\theta = 0$}\\
For $\theta =0$,
there is an energy gap $\Delta$, and the susceptibility
is simply that of a dilute, thermally activated, classical gas
of triplet magnons above the gap; these contribute an
exponentially small susceptibility~\cite{tsvelik,troyer}
\begin{equation}
\chi_u ( T) = \left( \frac{2 \Delta}{\pi T c^2} \right)^{1/2}
e^{-\Delta/T}~~~~~~~~~~;~~T < \Lambda_{\overline{MS}},~\theta = 0.
\label{e1}
\end{equation}
Experimentally, we can view (\ref{e1}) as the definition of the
gap $\Delta$ and the velocity $c$, which are to be determined by
fitting measurements to (\ref{e1}). The correlation length, $\xi (T)$ takes a
finite, $T$-independent value in this quantum paramagnet, up to correlations
exponentially small in $\Delta/T$; for the
case where $\Delta$ is significantly smaller than $J$, we
have~\cite{olaf,sudip}
\begin{equation}
\xi ( T) = \frac{c}{\Delta}~~~~~~~~~~;~~T < \Lambda_{\overline{MS}},~\theta = 0.
\label{e1a}
\end{equation}

\noindent
(A.II) \underline{$\theta = \pi$}\\
For the gapless case, $\theta = \pi$, there are excitations with non-zero
spin at arbitrarily low energies, and so $\chi_u (T)$ remains non-zero
as $T \rightarrow 0$~\cite{eggert,nomura}:
\begin{equation}
\chi_u (T) = \frac{1}{2 \pi c} \left[ 1 + \frac{1}{2 \ln (T_0/T)}
- \frac{\ln(\ln(T_0/T))}{4 \ln^2 (T_0 /T)} +
\ldots \right]~~~~;~~T < \Lambda_{\overline{MS}},~\theta = \pi.
\label{e2}
\end{equation}
Again, this experimentally defines $c$ and a new temperature scale
$T_0$ which determines the onset of a logarithmic correction to
the $T=0$ susceptibility due to the slow flow into the fixed point
at $g=g_c$. The correlation length of this critical paramagnet
now diverges as $T \rightarrow
0$~\cite{schulz,ssnmr,star1,star2,nomura2}:
\begin{equation}
\xi (T) = \frac{c}{\pi T}\left[ 1 + \frac{1}{2 \ln (T_0/T)}
- \frac{\ln(\ln(T_0/T))-1}{4 \ln^2 (T_0 /T)} +
\ldots \right]~~~~;~~T < \Lambda_{\overline{MS}},~\theta = \pi.
\label{e2b}
\end{equation}
The dynamical properties of quantum antiferromagnets in the low
temperature regime $T < \Lambda_{\overline{MS}}$ have been
discussed at length in Ref~\onlinecite{kedar} for the gapped case
($\theta = 0$) and in Refs~\onlinecite{schulz,ssnmr,star1,star2} for the gapless
case ($\theta=\pi$).

Note that both $\Delta$ and $T_0$ are energy scales characterizing
the flow {\em into\/} the strong-coupling region. These should therefore
be universally related to $\Lambda_{\overline{MS}}$ which is
the scale of flows {\em out of\/} the weak coupling region.
We will discuss the universal relation shortly, once we have
defined $\Lambda_{\overline{MS}}$ more precisely.

\noindent
(B)\underline{$ T> \Lambda_{\overline{MS}}$}\\
Let us now consider the regime $\Lambda_{\overline{MS}} < T < T_{\rm
max}^{(1)}$; the upper-bound is necessary to ensure the
continuum quantum theory still applies. The existence of this
intermediate temperature regime requires that $\Lambda_{\overline{MS}} <
T_{\rm max}^{(1)}$, a condition that is not well satisfied for
small $S$ and $p$, and so this regime almost certainly does not
exist for $p=1$ and $S=1/2,1$, but there is evidence that it is
present for $p=1$, $S=2$~\cite{kim}.
Here we are controlled by physics in the vicinity of the
$g=0$ fixed point, and it should be possible to treat quantum
fluctuations in a renormalized perturbation theory in $g$.
As discussed in Ref~\onlinecite{kedar},
a non-perturbative treatment of the thermal fluctuations is still
necessary, but this can be carried out exactly because of the low
spatial dimensionality. The result of such a calculation is~\cite{kedar}
\begin{equation}
\chi_u (T) = \frac{1}{3 \pi c} \left[ \ln\left( \frac{4 \pi e^{-1-\gamma}
T}{\Lambda_{\overline{MS}}} \right) + \ln\ln \frac{T}{\Lambda_{\overline{MS}}} +
\ldots \right] ~~~~;~~ \Lambda_{\overline{MS}}
< T < T_{\rm max}^{(1)},
\label{e2a}
\end{equation}
where $\gamma$ is Euler's constant.
This result can also be viewed as the precise experimental
definition of $\Lambda_{\overline{MS}}$. For completeness, we also
quote the result in this regime for the correlation length, $\xi
(T)$:
\begin{equation}
\xi (T) = \frac{c}{2 \pi T} \left[ \ln( \frac{4 \pi e^{-\gamma}
T}{\Lambda_{\overline{MS}}})  + \ln\ln \frac{T}{\Lambda_{\overline{MS}}} +
\ldots \right] ~~~~; ~~\Lambda_{\overline{MS}}
< T < T_{\rm max}^{(1)}.
\label{e3}
\end{equation}

We now give the promised universal
relationship between the energy scales characterizing the
weak ($\Lambda_{\overline{MS}}$) and
strong ($\Delta$, $T_0$) coupling regimes. This requires non-perturbative knowledge
of the renormalization group flows, and can only be obtained from
an analysis of the full thermodynamic Bethe-ansatz solution of the
quantum field theory (\ref{action}). For $\theta =0$, such an
analysis was carried out in Refs~\onlinecite{hasen}, and the
result is now well known:
\begin{equation}
\Lambda_{\overline{MS}} = \frac{e}{8} \Delta~~~~~~~~~~;~~\theta =0.
\label{e4}
\end{equation}
For the gapless case, $\theta = \pi$, we will present
a derivation of the required relationship in Appendix~\ref{bethe}, building
upon some recent results~\cite{fat,luky}; our result is:
\begin{equation}
\Lambda_{\overline{MS}} = \left(\frac{\pi}{2}\right)^{1/2} e^{3/4-\gamma}
T_0~~~~~~~~~~;~~\theta =\pi.
\label{e5}
\end{equation}

We are now finally in a position to state precisely the main
objective of this paper. We will describe the dynamical properties
of one-dimensional antiferromagnets in the intermediate
temperature regime $\Lambda_{\overline{MS}} < T < T_{\rm
max}^{(1)}$. We quickly note as an aside that most of our
results actually hold over a wider regime of temperatures
$\Lambda_{\overline{MS}} < T < T_{\rm
max}^{(2)}$, where we will define the discuss the origin
of $T_{\rm max}^{(2)} > T_{\rm max}^{(1)}$ below; for now we ignore
this point.
A common treatment is possible for the $\theta=0$ and $\theta= \pi$
in this regime, with the two cases differing only in the input
values of the static parameters $\chi_u (T)$ and $\xi(T)$
as defined by (\ref{e2a},\ref{e3},\ref{e4},\ref{e5}).
Moreover, as $\Lambda_{\overline{MS}}$ becomes exponentially small
for moderate values of $S$ or $p$, this regime can be quite wide
and should be readily observable experimentally. Indeed, there is
good evidence from recent measurements of static properties in quantum
Monte Carlo simulations~\cite{kim} that this universal intermediate
temperature regime exists even for $S=2$ spin chains.

The formulation of the dynamics properties for $\Lambda_{\overline{MS}} < T < T_{\rm
max}^{(1)}$ was already discussed in Ref.~\onlinecite{kedar}.
The key point~\cite{kedar} is to notice that the energy of a typical spin-wave
excitation, which is of order $c \xi^{-1}(T)$, is parametrically
smaller than $T$ when $\xi$ obeys (\ref{e3}). So the thermal occupation
number of these spin-wave modes is large:
\begin{equation}
\frac{1}{e^{c \xi^{-1}/T} - 1} \approx \frac{T}{c \xi^{-1}(T)} >
1.
\label{cep}
\end{equation}
The second expression in (\ref{cep}) is the classical
equipartition value, which indicates that the spin-wave
excitations may be treated
classically. The classical partition function controlling these
fluctuations can be deduced by demanding that its correlations
match with (\ref{e2a},\ref{e3}), while the dynamic equation of
motions following by replacing the quantum commutators with
Poisson brackets. In this manner, the problem reduces to the
effective classical phase-space partition function~\cite{kedar}
\begin{eqnarray}
&& {\cal Z}_C =  \int {\cal D} {\bf n} (x) {\cal D} {\bf L} (x)
\delta( {\bf n}^2 - 1) \delta( {\bf L} \cdot {\bf n}) \exp\left(-\frac{{\cal H}_C}{T}
\right) \nonumber \\
&& {\cal H}_C = \frac{1}{2} \int d x \left[ T\xi (T)
\left( \frac{d {\bf n} }{d x} \right)^2
+ \frac{1}{\chi_{u \perp} (T)} {\bf L}^2\right].
\label{class}
\end{eqnarray}
Here ${\bf n} (x)$ is a classical variable representing the
orientation of the antiferromagnetic order and ${\bf L} (x)$ is
its classical, canonically conjugate angular momentum. Because ${\bf n} (x)$
is of unit length, its motion is always in a direction orthogonal
to its instantaneous direction, and there is no radial kinetic
energy: the square of angular momentum represents the entire
kinetic energy, and $\chi_{u \perp}$ is the moment of inertia of
the fluctuating ${\bf n}$. The value of $\chi_{u, \perp}$ can be
determined by realizing~\cite{kedar} that ${\bf L} (x)$ is simply
the classical limit of the quantum operator corresponding to the
magnetization density
\begin{equation}
{\bf L} = -\frac{\delta {\cal L}}{\delta H};
\end{equation}
then
demanding that ${\cal Z}_C$ reproduce the correct uniform
susceptibility to a uniform external field under which
${\cal H}_C \rightarrow {\cal H}_C - \int dx \, {\bf H} \cdot {\bf
L}$, we obtain
\begin{equation}
\chi_u (T) = \frac{2}{3} \chi_{u\perp} (T);
\label{e6}
\end{equation}
the factor $2/3$ comes from the constraint ${\bf L} \cdot {\bf n} =
0$, so there are only two independent components of ${\bf L}$ at
each spatial point.

Notice that (\ref{class}) involves
a functional integral over
the commuting fields ${\bf n}$ and its conjugate momentum
${\bf L}$ as functions only of the spatial co-ordinate $x$, but is
independent of the real time, $t$. It therefore yields only equal-time
correlation functions, as is the usual situation in classical
statistical mechanics. To obtain unequal time correlators, we have
to separately specify the equations of motion, and these are
obtained by replacing quantum commutators with Poisson brackets.
For the fields ${\bf n} (x)$, ${\bf L} (x)$ these are
\begin{eqnarray}
\left\{ L_{\alpha} (x) , L_{\beta} (x') \right\}_{PB} &=&
\epsilon_{\alpha \beta \gamma} L_{\gamma} (x) \delta(x-x') \nonumber \\
\left\{ L_{\alpha} (x) , n_{\beta} (x') \right\}_{PB} &=&
\epsilon_{\alpha \beta \gamma} n_{\gamma} (x) \delta(x-x') \nonumber \\
\left\{ n_{\alpha} (x) , n_{\beta} (x') \right\}_{PB} &=& 0.
\label{e7}
\end{eqnarray}
where $\alpha,\beta\gamma = 1,2,3$.
The equations of motion (in real time) now follow from the Hamiltonian ${\cal
H}_C$, and they are
\begin{eqnarray}
\frac{\partial {\bf n}}{\partial t} &=& \frac{1}{\chi_{u \perp} (T)} {\bf L} \times {\bf n}
\nonumber \\
\frac{\partial {\bf L}}{\partial t} &=& (T \xi (T)) {\bf n} \times \frac{\partial^2 {\bf n}}{
\partial x^2}
\label{e8}
\end{eqnarray}
We are now interested in unequal time correlation functions of
(\ref{e8}), averaged over the classical ensemble of initial
conditions specified by ${\cal Z}_C$.

To complete the quantum-to-classical mapping, we
recall~\cite{kedar} the relationship between the correlations of
the underlying antiferromagnet and the quantum field theory ${\cal
Z}_Q$,
and those of the classical non-linear wave problem defined by
(\ref{class}) and (\ref{e8}). Correlations of the antiferromagnet
in the vicinity of the antiferromagnetic wavevector are given by
the correlations of ${\bf n} (x, t)$ under ${\cal Z}_Q$, and are
related to those in the classical problem by~\cite{kedar}
\begin{equation}
\langle {\bf n} (x, t) \cdot {\bf n} (0, 0) \rangle_{Q}
= {\cal A} \left[ \ln\left(\frac{T}{\Lambda_{\overline{MS}}}
\right) \right]^2 \langle {\bf n} (x, t) \cdot {\bf n} (0, 0)
\rangle_{C},
\label{e9}
\end{equation}
where the subscript $Q$ represents averages under the quantum
partition function ${\cal Z}_Q$, the subscript $C$ represents
averages under the classical dynamical problem defined by
(\ref{class},\ref{e8}), and ${\cal A}$ is an overall $T$-independent
normalization related to the amplitude of the correlations at
$T=0$. Next, correlations of the antiferromagnet
in the vicinity of zero wavevector are given by
the correlations of ${\bf L} (x, t)$ under ${\cal Z}_Q$, and are
essentially equal to those in the classical problem by~\cite{kedar}
\begin{equation}
\langle {\bf L} (x, t) \cdot {\bf L} (0, 0) \rangle_{Q}
=  \langle {\bf L} (x, t) \cdot {\bf L} (0, 0)
\rangle_{C}.
\label{e10}
\end{equation}
The absence of an overall rescaling factor here is related to the
conservation of the total magnetization density.

The main objective of this paper is to evaluate
$\langle {\bf n} (x, t) {\bf n} (0, 0)
\rangle_{C}$ and $\langle {\bf L} (x, t) {\bf L} (0, 0)
\rangle_{C}$. An important property of these correlators is that
they satisfy simple scaling laws which allow us to completely
scale away all dependencies on $\xi(T)$ and $\chi_{u \perp} (T)$,
and to express everything in terms of parameter-free, universal
functions. These scaling laws follow from the fact that
(\ref{class},\ref{e8}) define a continuum classical problem which
is free of all ultraviolet divergences: this will become evident
from our analytic computations in Section~\ref{short} and the
numerical results of Section~\ref{numerics}. Consequently, simple
engineering dimensional analysis involving rescaling of $x$, $t$
and ${\bf L}$
can be used to absorb dependencies on the dimensionful parameters.
In this manner, it is not difficult to show that~\cite{kedar}
\begin{eqnarray}
\langle {\bf n} (x, t) \cdot {\bf n} (0, 0)
\rangle_{C} &=& \Phi_{\bf n} \left( \overline{x}, \overline{t} \right) \nonumber
\\
\langle {\bf L} (x, t) \cdot {\bf L} (0, 0)
\rangle_{C} &=& \left( \frac{T \chi_{u \perp} (T)}{\xi (T)} \right)
\Phi_{\bf L} \left( \overline{x}, \overline{t} \right),
\label{e11}
\end{eqnarray}
where
\begin{eqnarray}
\overline{x} &\equiv& \frac{x}{\xi(T)}\nonumber \\
\overline{t} &\equiv& t \left[\frac{T}{\xi(T) \chi_{u \perp}
(T)}\right]^{1/2},
\label{e11a}
\end{eqnarray}
and $\Phi_{\bf n} ( \overline{x}, \overline{t} )$ and
$\Phi_{\bf L} ( \overline{x}, \overline{t} )$ are universal
scaling functions. It is the primary task of this paper to
determine these scaling functions. In principle, these scaling
functions are determined by solving (\ref{class},\ref{e8})
after setting all parameters equal to unity, $T=\xi=\chi_{u
\perp}=1$, while replacing $x$,$t$ by
$\overline{x}$,$\overline{t}$. Notice that the resulting equation
is then parameter-free, and so there is no explicit small parameter in
which any kind of expansion can be carried out; nothing short of
an exact solution will do. We think this reasoning invalidates
some of the conjectures on exactness of
results made in Ref~\cite{reiter}, as we
will discuss further in Section~\ref{expts}.
However, when the arguments of the scaling functions are
themselves small, {\em i.e.} $|\overline{x}|,|\overline{t}| \ll
1$, then a systematic perturbation expansion is possible, and this
will be presented in Section~\ref{short}.

We close this introductory discussion by returning to the issue of
the maximum temperature up to which these results can be applied
to lattice antiferromagnets. The appearance here of a classical
spin model suggests that one should think about classical spin
models obtained by starting directly from the lattice quantum spin
model, without the use of the quantum field theory ${\cal Z}_Q$ as
an intermediate step~\cite{nobert}. Such a classical description will only work
for large $S$, and we can ask the question of when the resulting
classical model can be described by a continuum classical theory,
which will clearly be the one defined by (\ref{class},\ref{e8}).
The correlation length of a classical spin antiferromagnet is of
order $J S^2 p a/T$, and a continuum description will work
provided this is larger than the lattice spacing $a$ and the width
of a ladder system $pa$. This gives us the estimate~\cite{nobert}
\begin{equation}
T_{\rm max}^{(2)} \sim  J S^2.
\label{e12}
\end{equation}
In the regime $T_{\rm max}^{(1)} < T < T_{\rm max}^{(2)}$,
we can use a purely classical description of the spin model: its
correlation length and uniform susceptibility will not be
universal, but has to computed for the specific model under
consideration. For the model with only nearest neighbor
exchange, a standard computation on the classical antiferromagnet
gives for the uniform susceptibility
\begin{equation}
\chi_u (T) = \frac{p}{6Ja}\left[ 1 + \left(1 - \frac{1}{p} \right)
\frac{J_{\perp}}{J} \right]^{-1}~~~~~~~~~~;~~T_{\rm max}^{(1)} < T <
T_{\rm max}^{(2)},
\label{e13}
\end{equation}
and for the
correlation length,
\begin{equation}
\xi (T) = \frac{JS^2 p a}{T}~~~~~~~~~~;~~T_{\rm max}^{(1)} < T <
T_{\rm max}^{(2)}.
\label{e14}
\end{equation}
It is satisfying to note that there is a precise agreement between
the  results (\ref{e2a},\ref{e3}) and (\ref{e13},\ref{e14}) at the
common
boundary of their respective regions of validity, $T \sim T_{\rm
max}^{(1)}$: for the nearest neighbor model under consideration
we use the estimates for $g$ and $c$ in (\ref{gest}), and then using
$\ln (T /\Lambda_{\overline{MS}}) \approx 2 \pi/g$, we find the
required agreement.

It appears useful to review the final status of the regimes of
validity of the model studied here. The universal, continuum, classical model
(\ref{class},\ref{e8}) describes all one-dimensional Heisenberg
antiferromagnets in the temperature regime $\Lambda_{\overline{MS}} < T < T_{\rm
max}^{(2)}$; this regime is wide and well defined for moderately large values of
$S$ or $p$, and there is evidence that it exists already for $p=1$, $S=2$~\cite{kim}.
The complete definition of this classical dynamical model requires
the input of the temperature-dependent static parameters $\xi (T)$ and $\chi_u
(T)$. In the regime $\Lambda_{\overline{MS}} < T < T_{\rm
max}^{(1)}$ these parameters are universally specified by
(\ref{e2a},\ref{e3}),
with $\Lambda_{\overline{MS}}$
given by (\ref{e4}) for gapped spin chains ($\theta=0$) and by
(\ref{e5}) for gapless spin chains ($\theta=\pi)$. In the higher
temperature regime $T_{\rm max}^{(1)} < T <
T_{\rm max}^{(2)}$, these parameters are given by (\ref{e13}) and
(\ref{e14}) for the model with only nearest neighbor exchange, and by related
non-universal expressions for other antiferromagnets.

The following sections contain technical details towards the
determination of the scaling functions $\Phi_{\bf n}$ and $\Phi_{\bf L}$
in (\ref{e11}), along with some theoretical analysis on the
relationship between integrability and diffusion.
We will begin in Section~\ref{short} by describing the analytical short time
expansion of the correlators $\Phi_{\bf n}$ and $\Phi_{\bf L}$.
The long time limit will then be studied numerically in
Section~\ref{numerics}. The subsequent Section~\ref{integ}
discusses issues which are somewhat peripheral to the main focus
of this paper: the continuum equations of motion (\ref{e8}) are known
to be integrable~\cite{faddeev}, and this raises numerous
fundamental questions on the nature of spin transport in
integrable systems. These will be addressed in Section~\ref{integ}
by the study of a simple, integrable, toy model whose spin
correlators can be determined in close form. Further, we will see that these
correlators have a striking similarity to those of (\ref{e8}).
Readers interested primarily in spin chains can omit
Section~\ref{integ} and skip ahead to
Section~\ref{expts} where we will describe the implications of our
results for experiments.

%%%%%%%%%%%%%%%%%%%%%%%%%%%%%%%%%%%%%%%%%%%%%%%%%%%%%%%%%%%%%%%%%%%%%%

\section{Short time expansion}
\label{short}

This section will determine the small $t$ expansion of the scaling
functions $\Phi_{\bf n}$ and $\Phi_{\bf L}$
in (\ref{e11}). Normally, there is a completely straightforward
way of determining the short time expansion of interacting
systems~\cite{forester}---it can be related, order-by-order, to
equal-time correlators involving higher moments of the fields.
However, this standard procedure does {\em not} work for the model
(\ref{class},\ref{e8}) of interest here. This is because
we are dealing with a continuum model with an
infinite number of degrees of freedom, ${\bf n} (x)$, ${\bf
L}(x)$, present at arbitrary short distance scales. If
we naively generate the moment expansion, we find that the terms
quickly acquire rather severe ultraviolet divergences.

A separate theoretical tool is necessary to generate the short
time expansion, and this shall be described here. We shall use an
analog of the field-theoretic method known as chiral perturbation
theory. As we shall see below, the expansion is actually in powers
of $|t|$---this implies a non-analyticity in the $t$-dependence at
$t=0$, which is in fact the reason for the ultraviolet divergences
in the moment expansion. The latter method only gives an analytic
expansion in $t$, by construction.

In this section, and in Appendix~\ref{details},
we will use units in which $T= \chi_{u \perp} (T) =
1$. However, we will retain explicit dependence on $\xi \equiv
\xi(T)$. It turns out to be quite useful to keep track of powers
of $\xi$. Indeed, our computations will be designed to generate an
expansion of the correlators in powers of $1/\xi$, and this is
a posteriori seen to be a short time expansion.
We will return to physical units in stating our final results.

We will
therefore consider the problem of unequal time correlation
functions of the non-linear partial differential equations
\begin{eqnarray}
\frac{\partial {\bf n}}{\partial t} &=&  {\bf L} \times {\bf n}
\nonumber \\
\frac{\partial {\bf L}}{\partial t} &=& \xi {\bf n} \times \frac{\partial^2 {\bf
n}}{
\partial x^2}
\label{short1}
\end{eqnarray}
when averaged over the ensemble of initial conditions defined by
the partition function
\begin{equation}
{\cal Z}_C =\int {\cal D} {\bf n} (x) {\cal D} {\bf L} (x) \delta(
{\bf n}^2 - 1) \delta( {\bf L} \cdot {\bf n}) \exp\left(-\frac{1}{2}
\int d x \left[ \xi \left( \frac{d {\bf n} }{d x} \right)^2+ {\bf
L}^2-2\xi m^2 n_z\right] \right)
\label{short2}
\end{equation}

The last term in the action represents a field of strength $\xi m^2$
turned on in the $z$ direction and serves as a regularization
parameter for our perturbation expansion.  At the end of our
calculations we shall let $m \rightarrow 0$.  From this partition
function, one can immediately find the equal time correlation
functions to be
\begin{eqnarray}
\langle {\bf L}(x,0)\cdot{\bf L}(0,0) \rangle &=& 2\delta (x) \nonumber\\
\langle {\bf n}(x,0)\cdot{\bf n}(0,0) \rangle &=& e^{-|x|/\xi};
\label{short5}
\end{eqnarray}
the subscript $C$ is implied on all averages in this section,
unless stated otherwise.

We first construct our perturbation expansion in powers of $1/\xi$
for the equal time
problem specified by (\ref{short2}), and check that we do arrive at the
correct correlation functions as specified in (\ref{short5}).  The
extension to the unequal time problem will then be straightforward.
First, the constraints on the fields ${\bf n}$ and ${\bf L}$ are
solved by introducing two complex scalar fields $\phi$ and $\psi$.
\begin{eqnarray}
n_x&=&\frac{1}{\sqrt{2 \xi}}(\phi+\phi^*) \nonumber \\
n_y&=&\frac{1}{i\sqrt{2 \xi}}(\phi-\phi^*) \nonumber \\
n_z&=&\sqrt{1-2 \phi \phi^* /\xi} \nonumber \\
L_x&=&\frac{1}{\sqrt{2}}(\psi+\psi^*) \nonumber \\
L_y&=&\frac{1}{i\sqrt{2}}(\psi-\psi^*) \nonumber \\
L_z&=&-\frac{\phi \psi^* + \phi^* \psi}{\sqrt{\xi}\sqrt{1-2 \phi
\phi^* /\xi}}
\label{short3}
\end{eqnarray}

We introduce this decomposition to the functional integral
(\ref{short2}) and expand the square roots in power series of $1/\xi$.
In this manner, we arrive at an interacting field theory with an infinite number
of interactions, with $1/\xi$ as the small coupling.  However, to any
particular order in perturbation theory in $1/\xi$, we only need to keep a finite set
of interactions terms.  We evaluate the correlation functions in
real space using the ordinary machinery of diagrammatic perturbation
theory.  It is well known that such a perturbation expansion is
plagued by infrared divergences.  The decomposition (\ref{short3})
takes it for granted that the ${\bf n}$ field is ordered in the $z$
direction and $\phi$ represents small spin wave fluctuations around
the ordered state.  The infrared divergences are a signature of the
fact that this assumption is wrong in one dimension.  By introducing
the external field $\xi m^2$, we introduce long range order into the
system and thus regularize the divergences.  The divergences show up
as poles in $1/m$.  But if we calculate the ${\rm O}(3)$ invariant
correlation functions as in (\ref{short5}), we find that all poles
cancel.  Keeping this in mind, it is not difficult to show that the
results of perturbation theory agree with (\ref{short5}) at every
order.  The details are to be found in Appendix~\ref{details}.

Once we have the correlation functions in the equal time ensemble, we
proceed to evaluate the unequal time correlation functions.  To do
that we again insert (\ref{short3}) into the equations of motion
(\ref{short1}) and expand in powers of $1/\xi$ to get
\begin{eqnarray}
\frac{\partial \psi }{\partial t} & = &
i \xi^{1/2} \left[
\frac{\partial^2 \phi}{\partial x^2}
%-m ^2\phi
+\frac{1}{\xi}\frac{\partial}{\partial
x}\left(\phi^2\frac{\partial\phi^*}{\partial x}\right) + \ldots
\right] \nonumber\\
\frac{\partial \phi }{\partial t} & = & -i \xi^{1/2} \left[ \psi +
\frac{1}{\xi}\left(\phi^2\psi^*\right) + \ldots \right] .
\label{short7}
\end{eqnarray}
Higher order
terms add further nonlinear interactions.  We solve the initial value problem
for each of the fields by an iterative strategy.  First the free wave
equation is solved and the solution is plugged back into the lowest
order nonlinear term to solve the problem to the first order.  To
evaluate the correlation functions, we just multiply the fields and
carry out the average over initial conditions.  The initial condition
averages are, of course, known from the calculations described above.
The technical details are relegated to Appendix~\ref{details}
and here we only
quote the results to one loop order:
\begin{eqnarray}
\left(\frac{\xi (T)}{T \chi_{u \perp} (T)} \right)
\langle {\bf L}(x,t) \cdot {\bf L}(0,0)\rangle= &&
\left(\delta(\overline{x}-\overline{t})+\delta(\overline{x}+\overline{t})\right)
\left(1-\frac{|\overline{t}|}{2}\right) \nonumber \\
&&~+\frac{1}{2}\left(\theta(\overline{x}+\overline{t})
-\theta(\overline{x}-\overline{t})\right) + {\cal
O}(\overline{x},\overline{t})
\label{short4}
\end{eqnarray}
\begin{eqnarray}
\langle {\bf n}(x,t) \cdot {\bf n}(0,0)\rangle_C = && 1-\frac{1}{2}(|\overline{x}+\overline{t}|
+|\overline{x}-\overline{t}|) \nonumber \\
&&~+\frac{1}{16}\left(3(\overline{x}+\overline{t})^2+3(\overline{x}-\overline{t})^2
+2|\overline{x}+\overline{t}||\overline{x}-\overline{t}| \right) +
{\cal
O}(\overline{x},\overline{t})^3 ,
\label{short6}
\end{eqnarray}
where $\overline{x}$ and $\overline{t}$ are defined in
(\ref{e11a}), and the delta function in (\ref{short4}) is
interpreted as ${\cal O}(\overline{x},\overline{t})^{-1}$.

In these results if we set $t=0$, we immediately recover the equal
time results (\ref{short5}) to the corresponding order in $1/\xi$.
The structure of the correlation function (\ref{short4}) reflects the
causal propagation of the conserved angular momentum ${\bf L}$.
The first term
simply represents the free propagation of
the angular momentum density which is completely concentrated on the
``light cone''.  However, interactions to order $1/\xi$ do modify the free
propagation and transfers the angular momentum density from the
surface of the light cone to its interior.  This is represented by the
second term which is non-vanishing only inside the light cone. It
can be checked that the spatial integral of (\ref{short4}) remains
independent of $t$, as must be the case due to conservation of
total angular momentum.

The result (\ref{short6}) has significant implications for the
dynamic structure factor, $S(k, \omega)$, of the antiferromagnetic
order parameter:
\begin{equation}
S(k, \omega) = \int dx dt \langle
{\bf n} (x,t) \cdot {\bf n}(0,0) \rangle_Q e^{-i(kx-\omega t)}.
\label{num1a}
\end{equation}
First, we note that the result (\ref{e9}) and the scaling form
(\ref{e11}), imply that the dynamic structure factor satisfies
\begin{equation}
S(k, \omega) = \left[ \frac{\xi(T)
\chi_{u \perp} (T)}{T} \right]^{1/2} S(k) \Phi_S (\overline{k},
\overline{\omega})
\label{num4}
\end{equation}
where $S(k)$ is the equal time structure factor, which is known
exactly (apart from the overall normalization ${\cal A}$)
\begin{eqnarray}
S(k) &=& \int_{-\infty}^{\infty} \frac{d \omega}{2 \pi} S(k, \omega)
\nonumber \\
&=& {\cal A} \left[ \ln \left(
\frac{T}{\Lambda_{\overline{MS}}} \right) \right]^2 \frac{2 \xi (T)}{1
+ k^2 \xi^2(T)},
\label{num5}
\end{eqnarray}
and $\Phi_S$ is a universal scaling function of
\begin{eqnarray}
\overline{k} &=& k \xi (T) \nonumber \\
\overline{\omega} &=& \omega \left[ \frac{\xi(T)
\chi_{u \perp} (T)}{T} \right]^{1/2},
\label{num6}
\end{eqnarray}
which describes the relaxation of the equal time correlations.
The prefactor in (\ref{num4}) has been chosen so that the
frequency integral over $\Phi_{S}$ is normalized to unity
for every $\overline{k}$
\begin{equation}
\int_{-\infty}^{\infty} \frac{d \overline{\omega}}{2 \pi} \Phi_S (
\overline{k},
\overline{\omega}) = 1.
\label{num7a}
\end{equation}
We will now show that, modulo some very mild assumptions, the
result (\ref{short6}) exactly fixes the form of $\Phi_S (\overline{k}, \overline{\omega})$
for $|\overline{k}| \gg 1$. In this short distance regime, we are
at distances shorter than the correlation length, and the
system should look almost ordered. So we may expect that the
spectrum consists of weakly damped spin waves, and this motivates
the following ansatz for $\Phi_S$ in the regime $|\overline{k}| \gg 1$
and $|\overline{\omega}| \sim |\overline{k}|$:
\begin{equation}
\Phi_S (\overline{k}, \overline{\omega}) = \frac{\gamma
(\overline{k})}{(\overline{\omega} - \overline{k})^2 + \gamma^2
(\overline{k})} + \frac{\gamma
(\overline{k})}{(\overline{\omega} + \overline{k})^2 + \gamma^2
(\overline{k})},
\label{short8}
\end{equation}
where $\gamma (\overline{k})$ is the unknown spin-wave damping
parameter; we will shortly determine the large $\overline{k}$
limit of $\gamma ( \overline{k})$.
We now have to take the
Fourier transform of (\ref{num4}), (\ref{num5}) and
(\ref{short8}),
compare the result with (\ref{short6}). In making this comparison,
we should keep in mind that the large $k, \omega$ behavior of $S(k, \omega)$
can only determine the {\em non-analytic terms} in the small $x,t$
expansion. First, the integral over frequencies can be performed
exactly for the form (\ref{short8}), and (\ref{num4}) implies
\begin{equation}
\langle {\bf n}(x,t) \cdot {\bf n}(0,0)\rangle_C =
\int \frac{ d \overline{k}}{ \pi} \frac{e^{i \overline{k}
\overline{x} - \gamma(\overline{k}) |\overline{t}| } \cos (\overline{k}
\overline{t})}{\overline{k}^2 + 1} + \ldots
\label{short9}
\end{equation}
We reiterate that (\ref{short8}) is valid only for $|\overline{k}| \gg
1$, and so only the $1/\overline{k}^2$ contribution from the $(\overline{k}^2 + 1)$
denominator in (\ref{short9})
can be taken seriously. Also, we already have a non-analytic $|\overline{t}|$
dependence in the exponential, and so we can expand this in
powers of $|\overline{t}|$; in this manner we reduce
(\ref{short9}) to
\begin{equation}
\langle {\bf n}(x,t) \cdot {\bf n}(0,0)\rangle_C =
\frac{2}{\pi} \int_{k_1}^{\infty} d \overline{k} \frac{ \cos( \overline{k}
\overline{x}) \cos (\overline{k}
\overline{t})}{\overline{k}^2} -
\frac{2 |\overline{t}|}{\pi} \int_{k_2}^{\infty} d \overline{k}
\frac{ \gamma (\overline{k}) \cos( \overline{k}
\overline{x}) \cos (\overline{k}
\overline{t})}{\overline{k}^2} + \ldots,
\label{short9b}
\end{equation}
where $k_{1,2}$ are some large positive constants, and we have
assumed that $\gamma$ is an even function of $\overline{k}$.
The values of the integrals over $\overline{k}$ surely depend upon
$p$ and $p'$, but the key observation is that the non-analytic
terms in $\overline{x}$ and $\overline{t}$ do not. This follows
from the result
\begin{equation}
\int_{k_1}^{\infty} \frac{dk}{k^{\nu}} \cos (kx) =  \frac{\pi |x|^{\nu-1}}{2 \Gamma(\nu)
\cos(\pi \nu/2)} +
\ldots,
\label{short9c}
\end{equation}
where all omitted terms can be written as a series in
non-negative, even integer powers of $x$ (this allows an additive
constant, independent of $x$). We now assume $\gamma (\overline{k} \rightarrow \infty)
\sim \overline{k}^{\alpha}$, and then demand consistency between
(\ref{short9b},\ref{short9c}) and the non-analytic terms in (\ref{short6}). It is easy to
see that we must have $\alpha = 0$, and so $\gamma (\infty)$ is a
constant. Further, the unknown additive constant associated with the second integral in
(\ref{short9b}) must be such that there is no single $|t|$ term in
the correlator. Applying (\ref{short9c}) to (\ref{short9b}) with
this understanding, we deduce that
\begin{equation}
\langle {\bf n}(x,t) \cdot {\bf n}(0,0)\rangle_C =
\ldots -\frac{1}{2}(|\overline{x}+\overline{t}|
+|\overline{x}-\overline{t}|)(1 - \gamma(\infty) |\overline{t}|)  +
\ldots,
\label{short10}
\end{equation}
where all omitted terms are either analytic in $\overline{x}$ and
$\overline{t}$, or involve subleading non-analyticities.
We should now compare (\ref{short10}) with (\ref{short6}) by
matching only the non-analytic terms in the vicinity of the light cone
$\overline{x} = \pm \overline{t}$. In this latter region
we can approximate $|\overline{x} + \overline{t}||\overline{x} - \overline{t}| $
in (\ref{short6}) by $2 |\overline{t}| |\overline{x} \pm
\overline{t}|$---then the non-analytic terms in (\ref{short10})
and (\ref{short6}) match perfectly, and we obtain one of our important exact results
\begin{equation}
\gamma (\infty) = \frac{1}{2}.
\label{short11}
\end{equation}
Returning to physical units via (\ref{num6}), we conclude that the
frequency $\Gamma (T) \equiv (\omega/\overline{\omega})
\gamma (\infty)$,
given by
\begin{equation}
\Gamma (T) = \frac{1}{2} \left[ \frac{T}{\xi(T)
\chi_{u \perp} (T)} \right]^{1/2},
\label{short12}
\end{equation}
describes damping of spin-waves for $|k| \xi(T) \gg 1$. We will discuss the experimental
implications of this result in Section~\ref{expts}.

Results for the spin-wave damping have been obtained earlier by
Reiter and Sj\"{o}lander~\cite{reiter}, in their studies of
classical lattice antiferromagnets. If we insert the classical values
(\ref{e13},\ref{e14}) into (\ref{short12}), we obtain
\begin{equation}
\Gamma (T) = \frac{T}{Sp}\left[ 1 + \left(1 - \frac{1}{p} \right)
\frac{J_{\perp}}{J} \right]^{1/2},
\label{short13}
\end{equation}
which agrees with their result for $p=1$.
Keep in mind, though, that our result (\ref{short12})
has a much wider regime of applicability, beyond temperatures in which
a purely classical thermodynamics holds. Further discussion on the
relationship between our and earlier results appears in Section~\ref{expts}.

%%%%%%%%%%%%%%%%%%%%%%%%%%%%%%%%%%%%%%%%%%%%%%%%%%%%%%%%%%%%%%%%%%%%%%

\section{Numerical Results}
\label{numerics}

The previous section allowed us to determine the two-point
correlators for small $|\overline{t}|$. Here we will present
numerical simulations which examine the large $|\overline{t}|$
limit. These were performed
on a discrete lattice realization of
(\ref{class},\ref{e8}), with lattice spacings of $\xi/16$ or
larger, and had only nearest neighbor couplings between the ${\bf n}$
vectors. Initial states were generated by thermalizing the system
by the Wolff algorithm~\cite{wolff}. The time evolution was
carried out by a fourth order predictor-corrector method, and its
accuracy was tested by keeping track of the conserved
total energy and the lengths of the ${\bf n}$ vectors.

Our simulations are similar to many earlier studies of classical
spin chains (see Refs~\cite{gerhard,george} and references therein).
However, there is an important difference in that we are
dealing with rotor variables ${\bf n}$, ${\bf L}$, rather
than classical spins ${\bf S}$ which obey Poisson bracket
relations like those for ${\bf L}$ in (\ref{e7}).

We will consider correlators of ${\bf n}$ and ${\bf L}$ in the following
two subsections.

\subsection{Correlations of ${\bf n}$}
\label{ncorr}

We first consider dynamic correlators of ${\bf n}$.
The aim of our simulations is to obtain results for the dynamic
structure factor, $S(k, \omega)$, in regimes beyond the
case $|\overline{k}| \gg 1$, $|\overline{\omega}| \sim |\overline{k}|$
which was studied by the short time expansion.
Our
results were obtained for two cases---at equal positions (local) and at
zero wavevector.

The local correlator is measured in NMR
experiments, and we computed the local dynamic structure factor, $S_l (\omega)$
defined by
\begin{eqnarray}
S_l (\omega) &=& \int_{-\infty}^{\infty} dt \langle
{\bf n} (0,t) \cdot {\bf n}(0,0) \rangle_Q e^{i\omega t} \nonumber \\
&=& \int_{-\infty}^{\infty} \frac{dk}{2 \pi} S(k, \omega),
\label{num1}
\end{eqnarray}
By (\ref{e9}) and (\ref{e11}), $S_l$ satisfies the scaling form
\begin{equation}
S_l (\omega) = {\cal A} \left[ \ln \left(
\frac{T}{\Lambda_{\overline{MS}}} \right) \right]^2 \left[ \frac{\xi(T)
\chi_{u \perp} (T)}{T} \right]^{1/2} \Phi_l ( \overline{\omega}),
\label{num2}
\end{equation}
where
$\Phi_l$ is a fully universal function (with no arbitrariness in its
overall amplitude or the scale of its argument) with a unit integral over
frequency
\begin{equation}
\int \frac{d\overline{\omega}}{2 \pi} \Phi_l (
\overline{\omega} ) = 1.
\label{num7}
\end{equation}

\begin{figure}
\epsfxsize=4in
\centerline{\epsffile{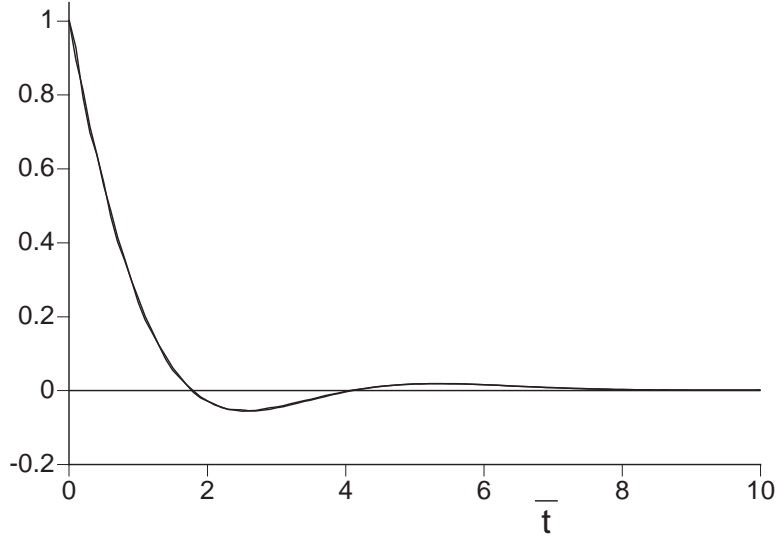}}
\caption{
The correlator $\langle {\bf n} (0,t) \cdot {\bf n} (0,0) \rangle_C$
as a function of $\overline{t}$, which is defined in
(\protect\ref{e11a}). Results with lattice spacing $\xi/16$ and $\xi/8$
are shown, and their near perfect overlap indicates we have
reached the continuum limit.
}
\label{fig2}
\end{figure}

\begin{figure}
\epsfxsize=4in
\centerline{\epsffile{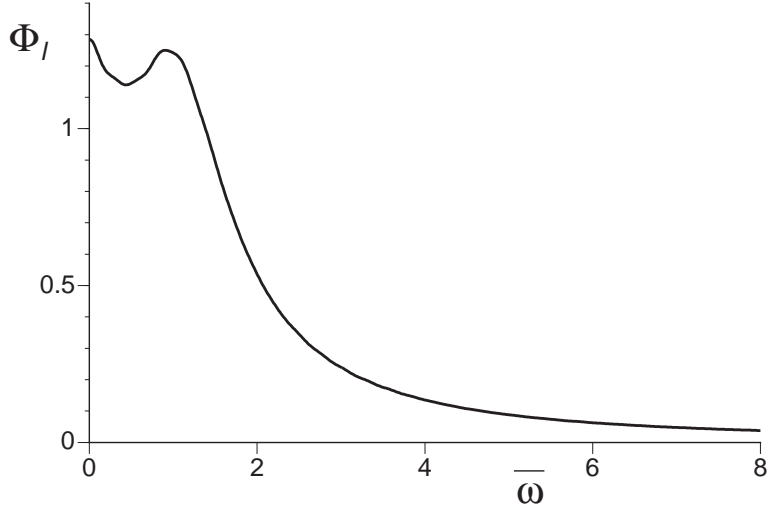}}
\caption{
The Fourier transform of Fig~\protect\ref{fig2} into frequency.
This yields the universal scaling function $\Phi_l$ defined in
(\protect\ref{num2}).
}
\label{fig3}
\end{figure}

Our results for the local time-dependent correlations are shown in
Fig~\ref{fig2} and its Fourier transform to frequency in
Fig~\ref{fig3}.
Two different lattice spacings were used, and the good overlap of
the data confirms that we are examining the continuum limit.
The correlations decay rapidly in time, but also show a brief, but
clear oscillation; this oscillation results in a finite frequency peak
in $S_l ( \omega) $.
We will discuss the physical origin of this
oscillation after we have considered the zero momentum correlator.

\begin{figure}
\epsfxsize=4in
\centerline{\epsffile{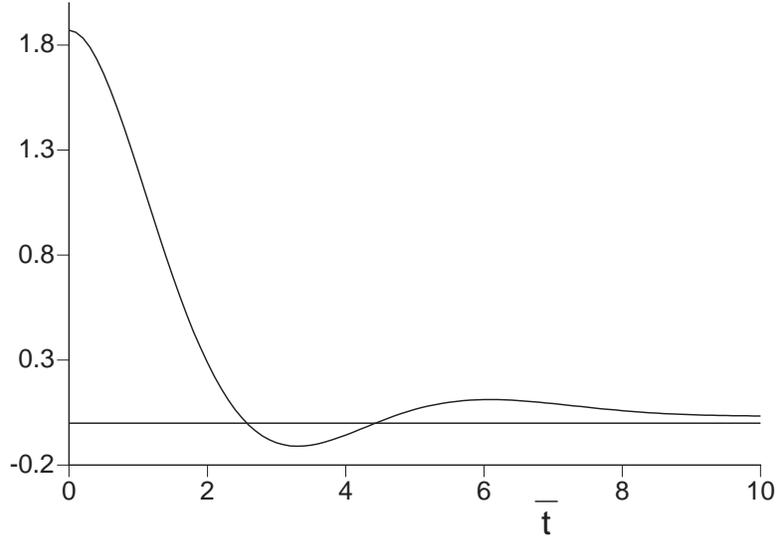}}
\caption{
The correlator $\int dx \langle {\bf n} (x,t) \cdot {\bf n} (0,0) \rangle_C /\xi(T)$
as a function of $\overline{t}$, which is defined in
(\protect\ref{e11a}). We used a lattice spacing $\xi/8$.
By (\protect\ref{num5}), the $t=0$ value of this
should be 2, and the difference is due to the finite lattice
spacing.
}
\label{fig4}
\end{figure}

\begin{figure}
\epsfxsize=4in
\centerline{\epsffile{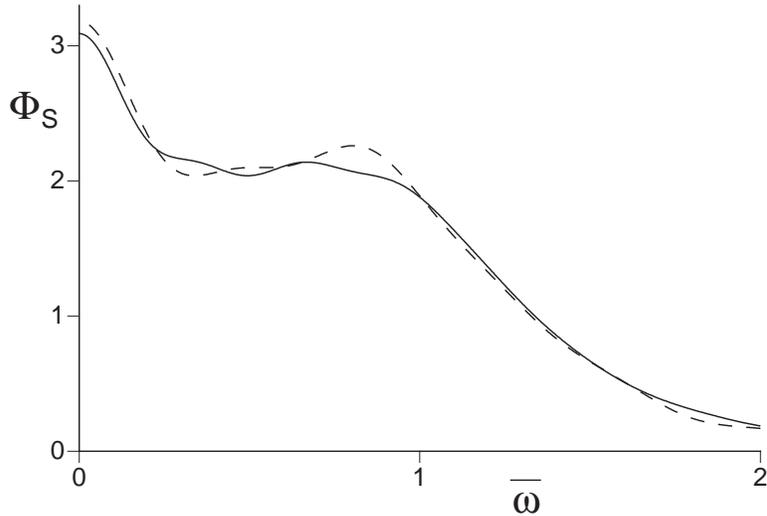}}
\caption{
The Fourier transform of Fig~\protect\ref{fig2} into frequency,
which when combined with (\protect\ref{num4}), leads to the
the universal scaling function $\Phi_S ( 0, \overline{\omega})$.
The full line is at lattice spacing $\xi/8$, and the dashed line
is for $\xi/16$. The data at $\xi/16$ is `noisier' because of insufficient
averaging for the longer time data.}
\label{fig5}
\end{figure}

Turning to the zero momentum correlator, we express our results in
the scaling form (\ref{num4}) and obtain values for the scaling
function $\Phi_S (0, \overline{\omega})$. We emphasize that the
ansatz (\ref{short8}) does {\em not} hold for $\overline{k} = 0$.
Our results for the $k=0$ correlators of ${\bf n}$ are shown in
the time domain in Fig~\ref{fig4} and after the Fourier transform to
frequency in Fig~\ref{fig5}.
As with the $x=0$ correlations above, the $k=0$ correlator shows a
rapid decay, along with a brief oscillation; the latter leads to a
finite frequency shoulder in $\Phi_S ( 0, \overline{\omega})$.

How do we understand this finite oscillation frequency observed in
both the $x=0$ and $k=0$ correlators of ${\bf n}$ ?
One way is to compare with the exactly known results~\cite{CSY,joli} of the model
with an $N$-component vector ${\bf n}$ in the limit of large $N$.
At $N=\infty$, $S(0,\omega)$ consists of a delta function
at a finite frequency $\omega \sim T /
\ln(T/\Lambda_{\overline{MS}})$. So we can view the finite
frequency as a remnant of the $N=\infty$ response at $N=3$.
However, there is a related, more physical, way to interpret it.
The underlying degree of freedom have a fixed amplitude, with $|{\bf n}| = 1$.
However,
correlations of ${\bf n}$ decay
exponentially on a length scale $\xi (T)$---so if we imagine coarse-graining
out to $\xi (T)$, it is reasonable to expect significant {\em amplitude
fluctuations}
in the coarse-grained field, which we call $\phi_{\alpha}$.
On a length scale of order $\xi (T)$,
we expect the effective potential controlling fluctuations of
$\phi_{\alpha}$ to have minimum at a non-zero value of
$|\phi_{\alpha}|$, but to also allow fluctuations in $|\phi_{\alpha}|$
about this minimum.
The finite frequency in Figs~\ref{fig3}, \ref{fig5} is due to the harmonic
oscillations of $\phi_{\alpha}$ about this potential minimum,
while the dominant peak at $\omega =0$ is due to angular
fluctuations along the zero energy
contour in the effective potential.
This is interpretation is also consistent with the large $N$
limit, in which we freely integrate over all components of ${\bf n}$,
and so angular and amplitude fluctuations are not
distinguished.

\subsection{Correlations of ${\bf L}$}
\label{lcorr}

We obtained numerical results only for the $x=0$ correlator of ${\bf
L}$. The short time behavior of this is given in (\ref{short4}).
At long times, we expect the conservation of total ${\bf L}$ to
be crucial in determining its asymptotic form. In particular, one
natural assumption is that the long time correlators of ${\cal L}$
are diffusive; in this case we expect
\begin{equation}
\langle {\bf L}(0,t)\cdot{\bf L}(0,0)\rangle=\frac{3 T \chi_u }{(4 \pi D t)^{1/2}}
\label{lcorr1}
\end{equation}
at large $t$. Consistency of this with the scaling form
(\ref{e11}), implies that the diffusion constant, $D$ must obey
\begin{equation}
D={\cal B}\frac{ T^{1/2} \left[ \xi (T) \right]^{3/2} }{
\left[ \chi_{u \perp} (T) \right]^{1/2}}
\label{lcorr2}
\end{equation}
where ${\cal B}$ is a dimensionless universal number.

Our numerical analysis of the autocorrelation was carried
out on a system of 800 sites. The predictor-corrector method
turns out to exactly conserve angular momentum and we maintained
energy conservation to 4 significant digits over the duration of
the simulation.
We averaged over
9,600 initial conditions.

\begin{figure}
\epsfxsize=4.0in
\centerline{\epsffile{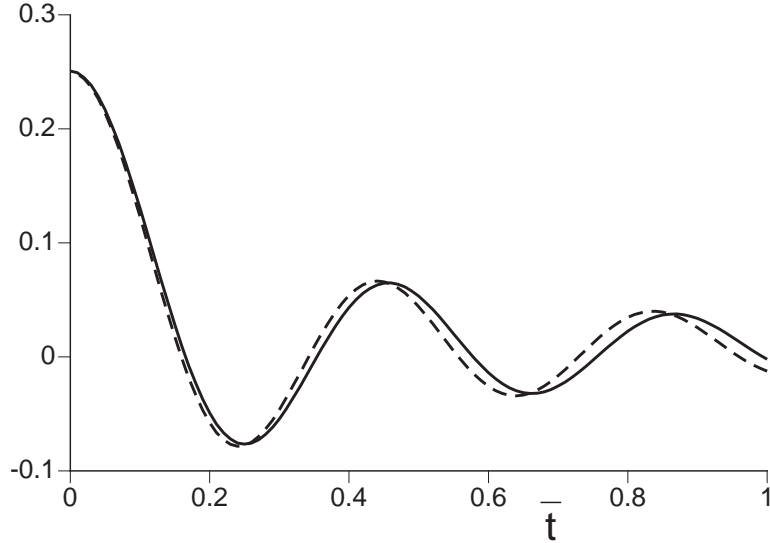}}
\caption{Numerical results (full line) for the
$(\xi(T)/T \chi_{u \perp} (T)) \langle {\bf L}(0,t)\cdot{\bf L}(0,0)\rangle$
correlation function for short times
$ \overline{t} < 1 $ on a lattice with spacing $\xi/8$.
The results are compared with short time
expansion (dashed line) in (\protect\ref{lcorr2a}),
valid for $\overline{t} \ll 1$.}
\label{fig6}
\end{figure}

First we tested our results against the known exact short time
expansion. This is shown in Fig~\ref{fig6}.
At these short times the lattice corrections are quite
significant, and our comparison in Fig~\ref{fig6} is with the
chiral perturbation theory carried out in the presence of a
lattice---the generalization of the result (\ref{short4}) to a
lattice model with nearest neighbor couplings and lattice spacing
$\epsilon \xi$ is
\begin{eqnarray}
&& \langle {\bf L} (x, t) \cdot {\bf L} (0,0) \rangle =
2 \int_{-\pi/\epsilon}^{\pi/\epsilon} \frac{dk}{2 \pi}
e^{i k \overline{x}} \cos (\omega_k \overline{t})
\left[ 1 + \int_{-\pi/\epsilon}^{\pi/\epsilon} \frac{dp}{2 \pi}
\frac{\cos(\omega_p \overline{t}) e^{i p \overline{x}} - 1}{\omega_p^2}
\right] \nonumber \\
&&~~~~~~~~~~+ 2 \left[ \int_{-\pi/\epsilon}^{\pi/\epsilon} \frac{dk}{2 \pi}
e^{i k \overline{x}} \sin (\omega_k \overline{t}) \right]^2 ,
\label{lcorr2a}
\end{eqnarray}
where $\omega_k = (2 ( 1 - \cos(\overline{k}
\epsilon))/\epsilon^2)^{1/2}$. It can be verified that
(\ref{lcorr2a}) reduces to (\ref{short4}) in the limit $\epsilon \rightarrow
0$.
As is clear from Fig~\ref{fig6}, the agreement between the analytical
and numerical computations is quite satisfactory.

\begin{figure}
\epsfxsize=4.0in
\centerline{\epsffile{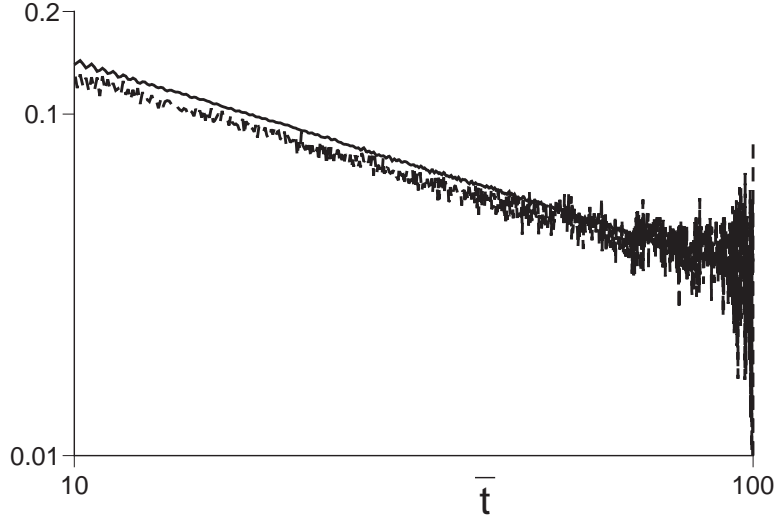}}
\caption{Numerical results for $(\xi(T)/T \chi_{u \perp} (T))
\langle {\bf L}(0 ,t)\cdot{\bf L}(0 ,0)\rangle$
as a function of $\overline{t}$ in a log-log plot.
The smoother line is for lattice spacing $\xi/8$, while
the noisier line is for lattice spacing $\xi/16$.
The agreement of the two results is evidence that this decay is a
property of the continuum limit.
A straight line
fit (not shown) to the $\xi/8$ data is almost perfect,
and its slope indicates that the
correlator decays as $\overline{t}^{-0.61}$.
An equally good fit to the data was obtained by the function
$a_1 /\protect\sqrt{\overline{t}} + a_2 / \overline{t}$, with the
second subleading term contributing only about a 10\% correction at the
largest $\overline{t}$.}
\label{fig7}
\end{figure}

Finally, we turn to the numerical results at large $\overline{t}$.
These are shown in Fig~\ref{fig7}.
A best fit to the data with a power law ${\overline{t}}^{-\alpha}$
gave an optimum value of $\alpha = 0.61$. However an equally
good fit to the data was obtained by the function
$a_1 /\protect\sqrt{\overline{t}} + a_2 / \overline{t}$, with the
second subleading term contributing only about a 10\% correction at the
largest $\overline{t}$ (we found $a_2/a_1 = 0.65$). This second fit is consistent with
diffusion---assuming this is the correct form, we obtain the
estimate for the numerical prefactor in (\ref{lcorr2}):
\begin{equation}
{\cal B} \approx 3.32.
\label{lcorr3}
\end{equation}

In their studies of the classical lattice model, Reiter and
Sj\"{o}lander~\cite{reiter}, also computed the spin diffusivity.
Diffusion is a property of the $|\overline{k}| \ll 1$ regime, and
we do not expect their perturbative techniques to be exactly
valid. Combining the classical values of $\chi_u (T)$, $\xi (T)$,
in (\ref{e13},\ref{e14}) with
(\ref{lcorr2}), their result translates into the value
${\cal B} = 1$. This value is clearly inconsistent with our
numerical result above.

%%%%%%%%%%%%%%%%%%%%%%%%%%%%%%%%%%%%%%%%%%%%%%%%%%%%%%%%%%%%%%%%%%%%%%

\section{Integrability and Diffusion}
\label{integ}

This section will examine an integrable toy model of
spin transport; readers interested mainly in the experimental implications
of our results
so far can move ahead to Section~\ref{expts}.
Others, not interested in the details of the toy model, may want
to jump to Section~\ref{discuss} where we will discuss general
implications of the toy model solution on spin transport in integrable
systems.

The toy model we shall introduce is a variant of an effective
model, considered in Ref~\onlinecite{kedar},
for the dynamics in the regime $T < \Delta$ for gapped chains.
Here, our strategy will be to introduce the model as worthy of
study in its own right, as it
is simple enough to allow determination of
the spin density correlator in closed form at all times.
In the long time limit, the correlator has a diffusive form, and so
this example {\em proves} that there is no general incompatibility
between integrability and diffusion. Further motivations in
examining this model are:\\
({\em i\/}) We shall show that the short time behavior of the toy
model is very similar to our result (\ref{short4}) for the
continuum wave model (\ref{class},\ref{e8}). This is suggestive,
and indicates that the long time diffusive behavior in
(\ref{lcorr1}) is not an unreasonable postulate.\\
({\em ii\/}) The correlators of the integrable toy model can also be studied
for the case of finite system of size $L$ with periodic boundary
conditions. This allows us to carefully examine the interplay of
the limits $t \rightarrow \infty$ and $L \rightarrow \infty$. The
diffusive form only appears if the $L \rightarrow \infty$ is taken
first. In the opposite order of limits we find recurrent behavior
with a great deal of structure dependent upon the microscopic
details of the model. We think this issue of the orders of limit
of $t \rightarrow \infty$ and $L \rightarrow \infty$ is
of considerable relevance to recent studies of $T>0$ transport in
integrable systems \cite{narozhny,zotos,narozhny2,mccoy,zotos2}, and this will be
discussed further in Section~\ref{discuss}.

\begin{figure}
\epsfxsize=2.0in
\centerline{\epsffile{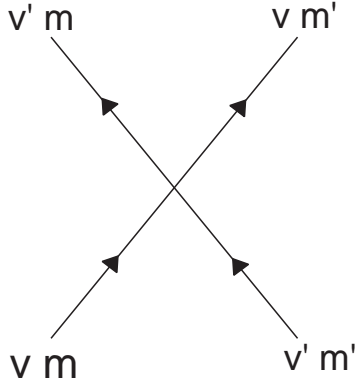}}
\caption{Collision of two particles. They initially have velocities
$v, v'$ and spins $m, m'$. After the collision they exchange velocities
but retain their spin.}
\label{fig8}
\end{figure}

We begin by describing the toy model.
Place $N$ point particles of equal mass at positions $x_i$ ($i=0,1,\ldots
N-1$) which are chosen independently from a uniform distribution
on a circle of length $L$. Now independently give each particle
a velocity $v_i$, drawn from some distribution $g(v)$, and a
`spin' $m_i$, drawn from some distribution $h(m)$. As the system
evolves, the particles will move in straight trajectories,
transporting their spin along with them. This will happen, until
two particles collide, and we now have to describe the nature of such collisions.
We will restrict the collisions to satisfy
the important constraints of conservation of  total energy,
momentum, and spin in each collision. The first two are already
sufficient to determine the fate of the velocities (see
Fig~\ref{fig8}): if we consistently label the particles from left
to right ({\em i.e.} as we move around the circle anti-clockwise, we always encounter
the particles in the order $x_0$, $x_1$, $x_2$, \ldots $x_{N-1}$), then
the particles will simply exchange velocities in each collision.
In other words, in a collision between particle $i$ and particle
$i+1$, the velocity of the particle $i$ after the collision is
that of particle $i+1$ before the collision, and vice versa.
How about the fate of the spins $m_i$ and $m_{i+1}$~? In principle,
we can
choose numerous possibilities interpolating between zero to total
reflection, consistent with conservation of total spin $\sum_{i}
m_i$. The exactly solvable models are the two extremes: zero or
total reflection. The case of zero reflection is rather trivial
and leads only to simple ballistic transport of spin along
straight lines. We will therefore consider only the case of total
reflection here: in the convention we are following of labeling
the particles here, this corresponds to the statement that
each $m_i$ is a constant of the motion (see Fig~\ref{fig8}).
So to summarize: in each collision the particles exchange
velocities but their spins `bounce off' each other.
We note that, although no exact solution is known,
we expect the long-time correlations of
a model with only partial reflection to be quite similar to
that of the total reflection case, but with renormalized transport
coefficients.

We shall be interested here in computing the correlators of the
`spin' density, $L(x,t)$, defined by
\begin{equation}
L(x,t) = \sum_{i=0}^{N-1} m_i \delta(x - x_i (t)),
\label{jepss1}
\end{equation}
where $x_i (t)$ are the positions of the particles: these consist
of piecewise straight lines which reflect at each collision.
We shall compute the two-point correlator of $L(x,t)$, averaged
over the ensemble of initial conditions defined above. Further, we
will choose our distributions $g(v)$ and $h(m)$ to be even
{\em i.e.} on the average, the total net momentum and spin are
zero. Then, because the initial momenta and spin are uncorrelated,
we have
\begin{eqnarray}
\langle L (x, t) L(0, 0) \rangle &=&
\sum_{i,i'} \langle m_i m_{i'} \rangle \langle
\delta(x-x_i (t)) \delta(x_{i'} (0) \rangle \nonumber \\
&=& \langle m^2 \rangle \sum_i \langle
\delta(x-x_i (t)) \delta(x_{i} (0) \rangle \nonumber \\
&=& \langle m^2 \rangle \rho P(x,t),
\label{jepss2}
\end{eqnarray}
where $P(x,t)$ is the probability that a particle at $x=0$ at
time $t=0$ is at the position $x$ at time $t$,
\begin{equation}
\langle m^2 \rangle = \sum_m m^2 h(m),
\label{jepss3}
\end{equation}
and $\rho = N/L$ is the density of particles.

Most of our results will be on a particular, simple, velocity
distribution, which is designed to mimic the properties of the
continuum model (\ref{class},\ref{e8}):
\begin{equation}
g(v) = \frac{1}{2} \left[
\delta(v-c) + \delta(v+c) \right].
\label{jepss4}
\end{equation}
So each particle is allowed to have only one of two velocities, $\pm
c$. This is similar to the fact that linear spin-waves in the
continuum wave model also have velocities $\pm c$.

The remainder of this section will describe the computation of the
function $P(x,t)$ using the method of Jepsen~\cite{jepsen}.
Although later more elegant solutions were put forward by Lebowitz and
Percus~\cite{lebowitz} for solving the
model in the thermodynamic limit, we shall use the relatively
cumbersome machinery of Jepsen because it allows us to consider finite
systems.

At time $t=0$ the $N$ particles are at random positions on the ring.
We shall put the origin of the
coordinate system at the location of particle 0.  The rest of the particles from $0,1,2,\cdots
N-1$ are numbered such that the $(i+1)$-th particle is immediately to
the right of the $i$-th one.  As a particle moves with uniform velocity, in a
``space-time'' diagram we can represent its motion as a straight line
which we'll call a trajectory.  When two trajectories cross, there is
a collision.  The particles bounce off each other, and in effect the
particles exchange trajectories.  So at the beginning the zeroth
particle starts on the zeroth trajectory and as this trajectory
crosses others, the zeroth particle moves onto a different trajectory.

Now define $A_{jk}(t)$ to be one if the particle $j$ is on the
trajectory $k$ at time $t$ and zero otherwise.  If we form an ensemble
of systems, the the average $\langle A_{jk}(t) \rangle$ is the
probability that the particle $j$ is on the trajectory $k$ at time $t$.
A knowledge of $A_{jk}(t)$ for all values of $j,k,t$ constitute a full
solution to the dynamics of the system.  The solution is defined by
\cite{jepsen}
\begin{eqnarray}
A_{jk}(t) & = & \frac{1}{N}\sum_{u} e^{-iju} \prod_{h=0}^{N-1}
S[u,w_{kh}]  \nonumber \\
u &  = &\frac{2 \pi l}{N}, \;\; l=0,1,2,\ldots, N-1; \;\; \sum_u \equiv \sum_{l=0}^{N-1}
\nonumber \\
S[u,w]& = & e^{inu}\;\; \mbox{when}\;\; (n-1)L < w \leq nL \;\;
\mbox{for each}\; n \nonumber\\
w_{kh}& = & x_k-x_h+(v_k-v_h)t .
\label{jep1}
\end{eqnarray}
We shall note some periodicity properties of
$S[u,w]$ and $A_{jk}(t)$ here.  By the above definition
\begin{equation}
S[u,w+L]=e^{iu}S[u,w],\;\; \mbox{hence}\;\; S[u,w+NL]=S[u,w].
\label{jep2}
\end{equation}
Also, for the distribution (\ref{jepss4}),
noting the fact that $|v_k-v_h|=0,2c$, we arrive at
\begin{equation}
A_{jk}\left(t+\frac{NL}{2c}\right)= \frac{1}{N}\sum_{u} e^{-iju}
\prod_{h=0}^{N-1} S \left[ u,w_{kh}+NL\frac{v_k-v_h}{2c} \right] =
A_{jk}(t).
\label{jep3}
\end{equation}
Let us define a time ${\cal T} \equiv L/c$ which is the time required by a
free particle to go once around the system.  Every trajectory returns
exactly to its starting point after this interval of time.  It is now
obvious that in a period of time $NL/c=N {\cal T}$, each particle will return to
its initial positions and velocities.  Thus the Poincare recurrence
time of the system, with the velocities chosen under
(\ref{jepss4}), is of order $N$, rather than being of the order of
$e^N$ or larger.

\subsection{Diffusion in the Thermodynamic Limit}

Now we go to the thermodynamic limit and address the question of
diffusion.  The limit is defined such that $N$ and $L$ approach
infinity while the density $N/L = \rho$ being held finite.
The zeroth
particle starts out at the origin and we ask what is the
probability, $P(y,t)$,
that it is at position $y$ at time $t$ (both $y$ and $t$ are held finite as
the limit $L \rightarrow \infty$ is taken).  This can be
written as
\begin{equation}
P(y,t)=\langle \delta(y-x_0(t)) \rangle=\sum_k\langle
\delta(y-x_k-v_kt)A_{0k}(t)\rangle,
\label{jep7}
\end{equation}
where the angular brackets an average over all possible initial
ensembles of velocity and positions of particles, while keeping the zeroth
particle at the origin.  The average can be evaluated exactly, and
there is a simple, closed form result:
\begin{eqnarray}
&& P(y,t) = \frac{1}{2}\left[ \delta(y+ct)+\delta(y-ct) \right]
e^{-\rho c |t|} \nonumber\\
&&~~ + \frac{\rho}{2}\left[ \theta(y+ct)-\theta(y-ct) \right]
e^{-\rho c |t|}
\left[\frac{c|t|}{\sqrt{c^2t^2-y^2}} I_1(\rho \sqrt{c^2t^2-y^2} )
+I_0(\rho \sqrt{c^2t^2-y^2})\right].
\label{jep4}
\end{eqnarray}
$I_0$ and $I_1$ are the modified Bessel functions of order zero and
one respectively.  The resemblance to the correlation function of the non-linear
wave model given
in (\ref{short4}) is clear. The first term in (\ref{jep4}) is a
delta function along the light cone, but its contribution
decreases exponentially with time. The second term lies within the
light cone, and becomes increasingly important for large time.
Also, if we take the short
time limit of (\ref{jep4}), we get
\begin{equation}
P(y, t) =  \frac{1}{2}\left[ \delta(y+ct)+\delta(y-ct) \right] (
1 - \rho c |t| + \ldots) + \frac{\rho}{2}\left[ \theta(y+ct)-\theta(y-ct)
\right],
\label{jepss5}
\end{equation}
which is precisely of the form (\ref{short4}). However, unlike
(\ref{short4}), we can now also study the long time limit
analytically. We take this limit within the light cone with
$y \sim \sqrt{t}$, and then the
asymptotic expansions of the modified Bessel functions yields
\begin{eqnarray}
P(y) &  \approx & \frac{1}{(4 \pi D t)^{1/2}}
\exp \left(-\frac{y^2}{4Dt} \right)\\
D & = & \frac{c}{2\rho},
\label{jepss6}
\end{eqnarray}
which is the diffusive form assumed for the classical wave model in
(\ref{lcorr1}).
As shown by Jepsen, this calculation can also be done for a
general velocity distribution $g(v)$, and provided the
distribution is symmetric in $v$, we obtain (\ref{jepss6}) but
with
\begin{equation}
D=\frac{1}{\rho}\int_{0}^{\infty}v g(v) d v.
\end{equation}

\subsection{Effect of Periodic Boundary Conditions in a Finite Geometry}

Here we go back to the finite system and think more about it.  The
fact that the Poincare recurrence time is only linear in $N$, is
because of the fact that the phase space becomes very restricted once
we allow only two possible velocities.  An interesting effect is that
the recurrence time can be very different depending upon whether there
are even or odd number of particles in the system.  The recurrence
time is the lowest common multiple of ${\cal T}$ (the time required for
trajectories to return to their initial position) and $NL/2c=N{\cal T}/2$ (the time
required for particles to come back to their initial trajectory).
With
an odd number of particles, $N=2p-1$, the recurrence time is
$N{\cal T}=(2p-1){\cal T}$ as mentioned above.  But if we add one single extra
particle to the system, the recurrence time almost becomes half
because $NL/2c=N{\cal T}/2=p{\cal T}$ is an exact multiple of ${\cal T}$.

In fact, the recurrence time could be even smaller.  If we take an
even number $N$ of particles and choose their velocities as $\pm c$
randomly, the most likely scenario is one where half of them have
velocity $+c$, and the rest have velocities $-c$.  In that case
\begin{equation}
A_{jk}\left(t+{\cal T}\right)= \frac{1}{N}\sum_{u} e^{-iju}
\prod_{h=0}^{N-1} S \left[ u,w_{kh}+ {\cal T} (v_k-v_h) \right]
\end{equation}
Amongst all the factors that we have in the product on RHS, half of
them will not contribute any phase to the product because for them
$v_k-v_h=0$.  The other half will contribute a phase of exactly
$\exp(\pm 2iu)$ each.  So the total phase contribution will be
$\exp(\pm 2iu N/2) \equiv 1$!  Hence the most probable recurrence time
for a random ensemble of even number of particles is ${\cal T}$.  It
turns out that with reflecting hard wall boundary conditions, the
recurrence time is $2{\cal T}$ regardless of the initial conditions.

Now let us see how the effect of the recurrence time might show up in
the probability distribution of a diffusing particle.  We shall
essentially try to find the autocorrelation function for a single
specific particle moving around in a ring for all times $t<{\cal
T}=L/c$ and $t>{\cal T}=L/c$.  We shall again choose this particle to
be the zeroth particle and at time $t=0$, its position and velocity
are $x_0\equiv 0,\,v_0$ respectively.  The probability distribution is
defined as before in (\ref{jep7}).  Since $A_{jk}(t)$ is periodic with
a period of $N{\cal T}/2$, all distribution functions will also be periodic
with the same period.  Moreover
\begin{eqnarray}
A_{jk}(N{\cal T}/2-t) & = & \frac{1}{N}\sum_{u} e^{-iju}
\prod_{h=0}^{N-1} S \left[ u,w_{kh}+(v_k-v_h)(N{\cal T}/2-t) \right]
\nonumber\\
& = &
\frac{1}{N}\sum_{u} e^{-iju}
\prod_{h=0}^{N-1} S \left[ u,w_{kh}-(v_k-v_h)t \right].
\end{eqnarray}
When we carry out the average over the initial conditions, the
particles $k$ and $h$ will have the velocities $+v_k,+v_h$ and
$-v_k,-v_h$ with equal probability.  So the probability distribution
function will satisfy
\begin{equation}
P(y,N{\cal T}/2-t)=P(y,t).
\end{equation}
Thus we need to evaluate this function only for $0 \leq t \leq N{\cal T}/4$.

\begin{figure}
\epsfxsize=4.0in
\centerline{\epsffile{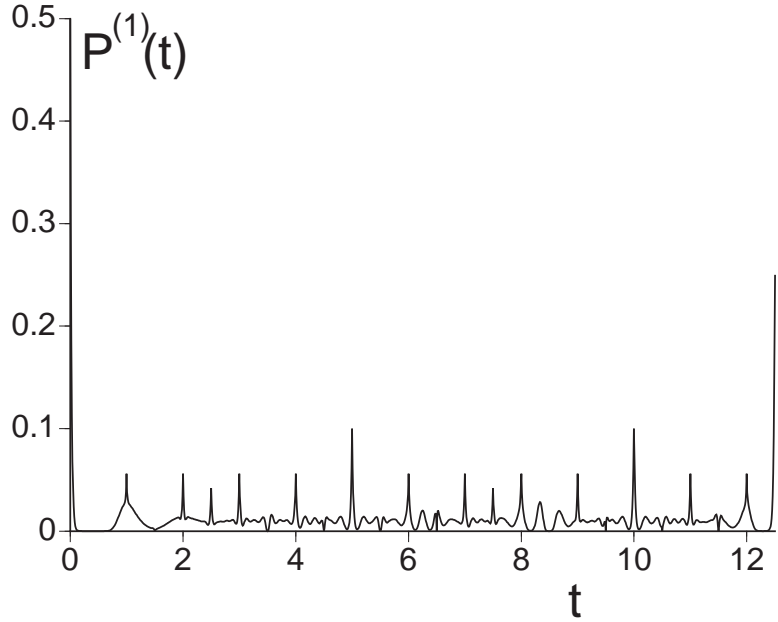}}
\caption{The probability distribution function $P^{(1)}(t)$ plotted as
a function of time $0 \leq t \leq N{\cal T}/4$ for $N=50$.  The unit of time is
${\cal T}$.  The stronger peaks at multiples of 5 is due to the fact that 5
is a prime factor of $N$}
\label{fig9}
\end{figure}

\begin{figure}
\epsfxsize=4.0in
\centerline{\epsffile{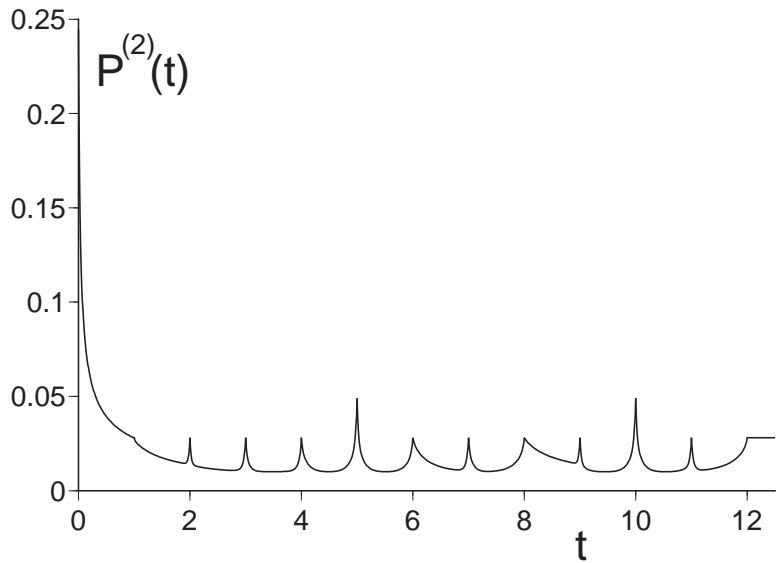}}
\caption{The probability distribution function $P^{(2)}(t)$ plotted as
a function of time for $N=50$ for $0<t<N{\cal T}/4$.  The unit of time is ${\cal T}$.}
\label{fig11}
\end{figure}

\begin{figure}
\epsfxsize=4.0in
\centerline{\epsffile{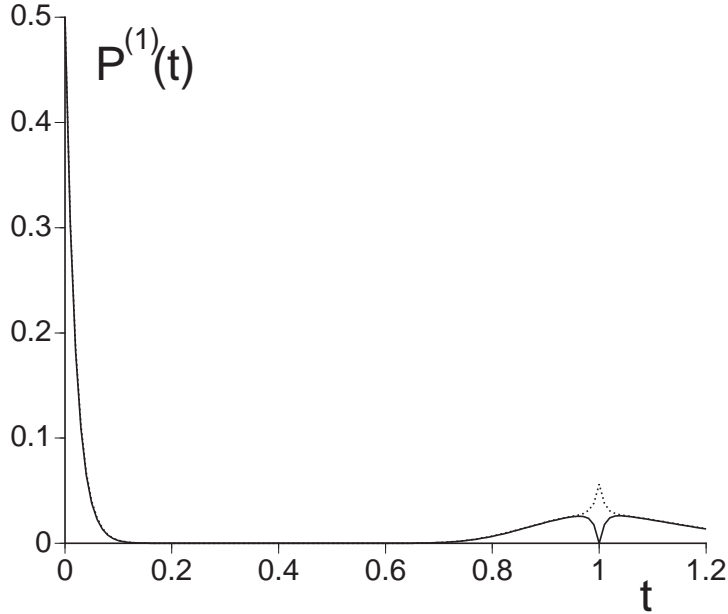}}
\caption{The probability distribution function $P^{(1)}(t)$ plotted as
a function of time for $N=50$ (dashed line) and $N=51$ (full line).
The unit of time is ${\cal T}$.
Notice the disappearance of the peak at $t=1$ for $N=51$.}
\label{fig10}
\end{figure}

The details of the evaluation of $P(y,t)$ are again relegated to the
appendix.  Let us represent the distribution function as
\begin{eqnarray}
P(y,t)=[\delta(y+ct)+\delta(y-ct)]P^{(1)}(t)+[\theta(y+ct)-\theta(y-ct)]P^{(2)}(y,t).
\end{eqnarray}
We write down explicit series solutions for $P^{(1)}(t)$ and
$P^{(2)}(0,t)$.  These sums could not be evaluated in a closed
analytic form.  So we carried out the sums numerically for specific values
of $N$ and $L$.  $P^{(1)}(t)$ and $P^{(2)}(0,t) $ are plotted in
Figs~\ref{fig9} and~\ref{fig11} respectively.
For times $t \ll {\cal T}$ and $N \gg 1$ with $N/L$ fixed, these
results reduce to the results (\ref{jep4}) derived in the
thermodynamic limit.  For $t>{\cal T}$, we see lots of complicated
structures.  Not all of them are well understood.  But some of the
prominent ones are easy to understand.  For example at times $t=n{\cal
T},\, n=1,2,3, \ldots$ we see peaks in $P^{(1)}$, each of whose height
turns out to be $1/\sqrt{2\pi N}$.  The origin of this is very
easy to understand.  Out of the whole ensemble of initial conditions,
a fraction of them will have exactly half of the particles with
velocity $+c$ and the others with velocity $-c$ (with the assumption
that $N$ is even).  Since these set of initial conditions have a
recurrence time of only ${\cal T}$, so they come back to the original
distribution after a time ${\cal T}$ and hence contribute to this
peak.  Using the binomial distribution, it is easy to prove that
the fraction of ensembles which have exactly half the particles with
one velocity and the other half with the opposite velocity is exactly
$\sqrt{2/\pi N}$.  The extra factor of $1/2$ comes from
the fact that we have both left and right going initial conditions for
our test particle.  If we change the number of particles by one, $N$
becomes odd.  So no set of initial conditions will have a recurrence
time ${\cal T}$, and these peaks will disappear as shown in
Fig~\ref{fig10}.

\subsection{Discussion}
\label{discuss}

Our study of an integrable toy model in this section has
highlighted the extreme importance of taking the limits of long time
$t \rightarrow \infty$, and large system size $L \rightarrow \infty$
with proper care.

If we send $L \rightarrow \infty$ {\em first},
then we explicitly demonstrated the existence of spin diffusion in
the subsequent long time limit. This is, of course, the correct
thermodynamic limit, and the presence of spin diffusion also makes
physical sense: once the limit $L \rightarrow \infty$ has been
taken, an infinite number of parameters are needed to specify the
initial thermal state of the system---diffusion then arises when
some local degree of freedom starts to sample an increasing number of these infinite
number of random initial conditions with the passage of time.

On the other hand, very different results were obtained for the
limit $t \rightarrow \infty$ at any fixed $L$. Here the integrable
nature of the system was immediately evident, and we observed a
bizarre set of recurrences dependent sensitively on details of the
microscopic Hamiltonian. The Poincare recurrence time of our toy
model was quite short, and this was clearly due to its
integrability. No sign of spin diffusion was seen.

A number of recent studies have examined the issue of $T>0$ spin
transport in integrable quantum
systems~\cite{narozhny,zotos,narozhny2,mccoy,zotos2}.
A model of particular
interest has been the $S=1/2$ XXZ chain. At the $SU(2)$ symmetric
point (the XXX chain), the low $T$ properties of this
antiferromagnet are expected to be in the universality class of
${\cal L}$ in (\ref{action}) at $\theta = \pi$. So studies of the
XXX model will explicate the nature of spin transport at temperatures
$T < T_0$ at $\theta = \pi$. This is a regime for which our paper
has no results (although, spin transport in the low $T < \Delta$ regime for $\theta  =0$
was studied in Ref~\cite{kedar}). However, we have examined the
higher temperature regime $T_0 < T < T_{\rm max}^{(2)}$ in
Section~\ref{lcorr}, and found that our numerical results are
not inconsistent with the presence of spin diffusion. It would then seem natural that
diffusion may also exist for $T < T_0$, although, this is, of
course, not a rigorous argument. The latest numerical evidence for
the XXX chain~\cite{narozhny2,zotos2} seems to be consistent with
the existence of diffusion.

Here ,we wish to issue a small caution towards the method used to
study transport in Refs~\cite{narozhny,zotos,narozhny2,zotos2}
(this caution does not apply to Ref~\cite{mccoy}). These works
computed a `stiffness', which is the co-efficient of zero frequency
delta function in the frequency dependent conductivity. By its very
construction, such a quantity is defined at $\omega = 0$ in a finite
system; so implicitly, the limit $\omega \rightarrow 0$ has been
taken before the $L \rightarrow \infty$ limit. The considerations
of this section make it amply clear that such a procedure is
potentially dangerous.

\section{Implications for experiments}
\label{expts}

First, we summarize the main theoretical results of this paper. We
have shown that there is an intermediate temperature range over which
the static and dynamic properties of
a large class of one-dimensional Heisenberg
antiferromagnets are described by the deterministic, continuum
model defined by (\ref{class}) and (\ref{e8}). For $p$-leg ladders
of spin $S$ ions, this temperature range rapidly becomes quite
wide as $S p$ increases. For $2 S p$ even, our universal results
hold for $\Delta < T < T_{\rm max}^{(2)}$, where $\Delta$ is the
ground state energy gap, and $T_{\rm max}^{(2)}$ is estimated in
(\ref{e12}), while for $2Sp$ odd, they hold for $T_0 < T < T_{\rm
max}^{(2)}$, where $T_0$ is an energy scale measuring the strength
of logarithmic corrections at the lowest $T$. Both $T_0$ and $\Delta$
become exponentially small as $S p$ increases, and so the
intermediate temperature regime is clearly defined. The dynamical
properties of such antiferromagnets are encapsulated in the
scaling forms (\ref{e11}), which relate them to universal
functions dependent only upon two thermodynamic parameters:
the antiferromagnetic correlation length $\xi (T)$,
and the uniform spin susceptibility $\chi_u (T)$. We obtained
information on the universal functions in Sections~\ref{short}
and~\ref{numerics}, including exact results on the spin-wave damping,
while exact results for $\xi(T)$ and $\chi_u (T)$
were presented in Section~\ref{intro}.

We now briefly review the work in the 70's on the dynamics of
classical antiferromagnetic chain
(Refs~\cite{blume,reiter} and references therein). As we saw in
Section~\ref{intro}, there is a window of temperatures
$  T_{\rm max}^{(1)} < T <  T_{\rm max}^{(2)}$ ($T_{\rm max}^{(1)}$ was estimated
in (\ref{t1est})), over which
(\ref{e13}) and (\ref{e14}) are valid, where our results apply to
purely classical models; so there is a common regime of
validity between our and earlier work.
These earlier classical results were all obtained on
studies of lattice antiferromagnets, and all used some variant of
the short-time moment expansion to extrapolate to the long time
limit by a physically motivated ansatz {\em e.g.\/} the memory
function formalism; however, there is a degree of arbitrariness in any
such ansatz. In a regime where their correlation length
$\xi \gg a$ (where $a$ is a lattice spacing), and the wavevectors $k a \ll
1$, their results should be described the continuum model we have
discussed here. However, we expect their short-time methods to be
exact only for $k \xi \gg 1$. Indeed, our paper is
the first to provide a proper description of the scaling structure, along
with quantitative information on the scaling functions, in the
non-perturbative regime $k \xi \ll 1$. Consistent with these
expectations, we saw in Section~\ref{short} that our result (\ref{short13}) for
the spin-wave damping for $k \xi \gg 1$, and $  T_{\rm max}^{(1)} < T <  T_{\rm max}^{(2)}$,
was in precise agreement
with that of Reiter and Sj\"{o}lander~\cite{reiter}. It should be
noted, however, that our result (\ref{short12}) for the damping
remains exact over a much wider
window of temperatures ($T_0, \Delta < T < T_{\rm max}^{(2)}$),
including when $\chi_u$ and $\xi$
have the quantum renormalized $T$-dependence in (\ref{e2a}) and
(\ref{e3}). Moreover, we believe, despite conjectures by
Reiter and Sj\"{o}lander to the contrary, that the results in
Ref~\cite{reiter} are not exact for $k \xi \ll 1$---we saw in
Section~\ref{lcorr} that our value for the spin diffusion
constant (assuming the existence of diffusion),
$D$, disagreed with theirs.

Turning to experiments, single chain ($p=1$) antiferromagnets with
$S>1$ which have been studied are $({\rm CD}_3)_4 {\rm N Mn Cl}_3$
(TMMC)~\cite{birg,hutchings} which has $S=5/2$, $({\rm C}_{10} {\rm H}_8 {\rm N}_2)
{\rm Mn Cl}_3$~\cite{granroth} which has $S=2$, and ${\rm Cs V
Cl}_3$~\cite{itoh} which has $S=3/2$.
We think it would be worthwhile to re-examine these
materials from a modern perspective, given the numerous exact results
that are now available.

Among static properties, neutron scattering experiments~\cite{birg} have
measured the correlation length, $\xi (T)$, and these have been
compared to purely classical theories in which $\xi(T)$ behaves like
(\ref{e14}). At lower temperatures, $\xi (T)$ should
exhibit the logarithmic temperature dependence in (\ref{e3}),
arising from quantum fluctuations. Combined with measurements of
the uniform susceptibility, $\chi_u
(T)$, a rather precise test of the quantum-renormalized
static theory should then be possible.

Dynamic tests of the theory have focused mainly on the linewidth
of the spin-wave excitations in the regime $k \xi \gg 1$.
The measured linewidths have been compared~\cite{hutchings,itoh}
with the prediction of the classical theory~\cite{reiter},
which yields the result (\ref{short13}). This is in general agreement with
the theory, but a quantitative discrepancy was
observed~\cite{itoh} for $S=3/2$.
We think it would be useful to compare the experiments with our
new exact result (\ref{short12}), while using the actual
experimentally observed values of $\xi$ and $\chi_u$.

We think future neutron scattering experiments should also examine the interesting
regime $k \xi \ll 1$. Here we have provided new, (numerically)
exact results in Section~\ref{numerics}. In particular, there is
some interesting physics in the structure of the
frequency-dependent lineshapes in Figs~\ref{fig3} and~\ref{fig5},
and these should be subjected to experimental tests.
Also, we can easily generate additional universal spectra at other positions
in the energy-momentum space, as needed.

The results in Section~\ref{numerics} also provide quantitative
predictions for NMR experiments on spin chains. The nuclear
relaxation rate $1/T_1$ is given by local, low frequency, dynamic
structure factor of the electronic spins. This has two
contributions, one from the ferromagnetic component given by
the correlator of ${\bf L}$, and the other from the
antiferromagnetic component given by the correlator of ${\bf n}$.
Let us parameterize the electronic spin ${\bf S}_i$ by
\begin{equation}
{\bf S}_i = (-1)^i S {\bf n} ( x_i) + \frac{a}{p} {\bf L} (x_i)
\label{expt0}
\end{equation}
where $a$ is the lattice spacing. If we assume that
\begin{equation}
\frac{1}{T_1} = \frac{\Gamma}{2}
\int_{-\infty}^{\infty} dt \, e^{i \omega_N t} \left\langle
(S_{xi}(t) + i S_{yi}(t))(S_{xi}(0) - i S_{yi}(0)) \right\rangle,
\label{expt0a}
\end{equation}
where $\Gamma$ is related to the hyperfine coupling, and $\omega_N \rightarrow 0$
is the nuclear Larmor frequency. The electron spin correlator has
to be evaluated in the presence of an applied magnetic field $H$,
and the electron Larmor precession can usually be neglected.
However, for the case where there is spin diffusion, as in the
assumed form (\ref{lcorr1}), then this electron precession must be
included for the Fourier transform is divergent at low
frequencies. Combining (\ref{expt0}) and (\ref{expt0a}) with the
results of Section~\ref{numerics}, we obtain
\begin{equation}
\frac{1}{T_1} = \Gamma \left( \frac{{\cal A} S^2}{3} \left[ \ln \left(
\frac{T}{\Lambda_{\overline{MS}}} \right) \right]^2 \left[ \frac{\xi(T)
\chi_{u \perp} (T)}{T} \right]^{1/2} \Phi_l (0) + \frac{T \chi_u (T) (a/p)^2 }{\sqrt{2 D H}}
\right),
\label{expt1}
\end{equation}
where $D$ is estimated
in (\ref{lcorr2},\ref{lcorr3}). If we ignore logarithmic factors,
the first antiferromagnetic term in (\ref{expt1}) is of order $\Gamma/T$
while the second ferromagnetic term is of order $(\Gamma/T) (Ta/c)^2
(T/H)^{1/2}$; either term could be dominant, depending upon the
magnitude of $H$. Further, our result (\ref{expt1}) has assumed
the existence of spin diffusion, but we expect that (\ref{expt1})
will provide a reasonable quantitative
estimate of the $H$ and $T$ dependence
for experimental purposes, even if this assumption is not
entirely correct in its details: there is clearly a long-time
tail in Fig~\ref{fig7}, even if it is not precisely diffusive.

\begin{figure}
\epsfxsize=4.0in
\centerline{\epsffile{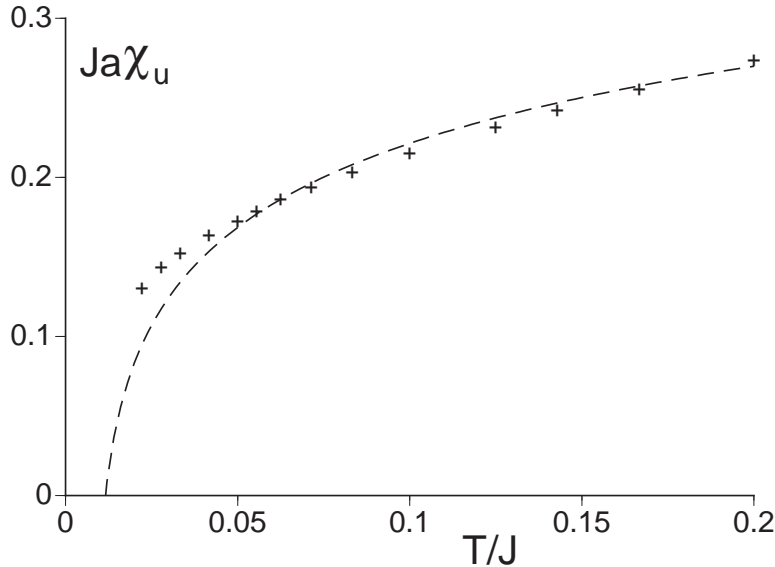}}
\caption{
Comparison of the numerical results of Ref.~\protect\cite{beat}
(plus marks) for the uniform susceptibility, $\chi_u$, of
a 5-leg ladder with $S=1/2$ with the theoretical predictions of
 (\protect\ref{e2a},\protect\ref{e5}) (dashed line).
All exchange constants are nearest neighbor and have magnitude
$J$, and the lattice spacing is $a$. The value of $\chi_u$ is {\em
per rung.}}
\label{fig13}
\end{figure}

Finally, we compare our theoretical predictions with quantum Monte
Carlo simulations on odd leg ladders.
Numerical results for $\chi_u (T)$ have been obtained recently by
Frischmuth {\em et al.}~\cite{beat} on antiferromagnets with $S=1/2$
and $p=3,5$. We compared their results for $p=5$, with our new result (\ref{e2a},\ref{e5}):
this is shown in  Fig~\ref{fig13}.
The fitting parameters in this comparison are the values of $c$ and
$T_0$. There is an arbitrariness in choosing the ranges of $T$
over which to fit the
intermediate $T$ prediction (\ref{e2a}), and this can lead to some
variation in the values of $c$ and $T_0$.  A reasonable
set of values are $c = 2.06 Ja$ and $T_0 = 0.0058 J$, are used in
Fig~\ref{fig13}. The value of $c$ is
roughly consistent with that estimated earlier~\cite{beat},
but the value of $T_0$ does appear to be rather small.

\acknowledgements

We thank I.~Affleck, K.~Damle, B.~Frischmuth, Y.-J.~Kim, B.~McCoy,
G.~M\"{u}ller, F.~Naef,
B.~Narozhny, G.~Reiter,
T.M.~Rice, M.~Takigawa and M.~Troyer for helpful
discussions.
This research was supported by NSF Grant No DMR 96--23181.

\appendix

\section{Crossover energy scales for $\theta = \pi$}
\label{bethe}
This appendix will derive the relationship (\ref{e5}) between the
two energy scales $T_0$ and $\Lambda_{\overline{MS}}$ associated with
the quantum ${\rm O} (3)$ non-linear sigma model
at $\theta = \pi$. This model has a flow from
the high energy fixed point at $g=0$ to the low-energy fixed point
at $g=g_c$. The flows near both fixed points are marginal: $\Lambda_{\overline{MS}}$
is the energy scale characterizing the flow away from the the $g=0$
fixed point, while $T_0$ is the scale characterizing the flow into
the $g=g_c$ fixed point. These scales appear in the logarithmic
corrections that appear in both the high temperature (Eqns
(\ref{e2a},\ref{e3})) and low temperature (Eqn (\ref{e2})) limit.
A complete Bethe ansatz analysis of the flow between the two fixed
points matches the two scales, and leads to the relationship
(\ref{e5}).

We will perform the matching by considering various limiting
regimes of the free energy density, ${\cal F}_Q$, of ${\cal Z}_Q$ as a function of $T$
and $H$. This is will allow us to make an intricate series of mappings between
numerous results which have appeared recently in the literature.

First, let us consider the low $T$ and low $H$ regime, where
$T, H \ll T_0$. In this regime, the model is in the vicinity of
the $g=g_c$ fixed point, which is the $k=1$, $SU(2)$
Wess-Zumino-Witten model. We know from (\ref{e2}), that for
$H \ll T \ll T_0$, the free energy has a contribution
\begin{equation}
{\cal F}_Q  = - \frac{H^2}{4 \pi c} \left[1 + \frac{1}{2 \ln (T_0
/T)} \right] + \ldots~~~~~; H \ll T \ll T_0.
\label{app1}
\end{equation}
From this we anticipate that for $T \ll H \ll T_0$, we will have a
corresponding contribution
\begin{equation}
{\cal F}_Q  = - \frac{H^2}{4 \pi c} \left[1 + \frac{1}{2 \ln (C_1 T_0
/H)} \right] + \ldots~~~~~; T \ll H \ll T_0,
\label{app2}
\end{equation}
where $C_1$ is a universal number we would like to determine. The
universality of $C_1$ implies that we can use any model which is
in the vicinity of the $k=1$, $SU(2)$
Wess-Zumino-Witten fixed point. In particular, Lukyanov~\cite{luky} has
recently computed the detailed $H$ and $T$ dependence of the free
energy of the $S=1/2$ antiferromagnetic chain with nearest neighbor exchange
which fulfills this requirement,
and we can use his results to obtain $C_1$. In particular, from (3.18)
and (3.20) of Ref~\onlinecite{luky} we determine that
\begin{equation}
T_0 = \left(\frac{\pi}{2} \right)^{1/2} e^{\gamma + 1/4} J
\label{app3}
\end{equation}
for the nearest neighbor $S=1/2$ antiferromagnet, and that
\begin{equation}
{\cal F}_Q  = - \frac{H^2}{4 \pi c} \left[1 + \frac{1}{2 \ln (2 \pi e^{-\gamma} T_0
/H)} \right] + \ldots~~~~~;~ H \ll T_0,~~T=0.
\label{app4}
\end{equation}
We emphasize that the result (\ref{app3}) is non-universal, while
(\ref{app4}) is universal {\em i.e.} only the latter is a property
of the continuum ${\rm O} (3)$ non-linear sigma model at $\theta =
\pi$. In the result (\ref{app4}) we have set $T=0$, as this is the
limit in which we shall use in the following.

Let us now consider the vicinity of the $g=0$ fixed point for
non-zero $H$ at zero temperature {\em i.e.\/} for $H \gg \Lambda_{\overline{MS}} \sim T_0$
at $T=0$.  In this case, a renormalized perturbation
theory in $g$ can be used to determine the free energy density, and this
was carried out by Hasenfratz {\em et al}~\cite{hasen}.
They obtained
\begin{equation}
{\cal F}_Q = - \frac{H^2}{4 \pi c} \ln \left( \frac{H}{e^{1/2}
\Lambda_{\overline{MS}}} \right)+ \ldots~~~~~;~H \gg
\Lambda_{\overline{MS}},~~T=0
\label{app5}
\end{equation}

Finally, we need to match (\ref{app4}) and (\ref{app5}) by
computing the $T=0$ free energy for $H \neq 0$ in both the low and
high field limits of the ${\rm O}(3)$ non-linear sigma model at
$\theta=\pi$. Fortunately, precisely this computation was carried
out by Fateev, Onofri and Zamolodchikov~\cite{fat}. They obtained
(\ref{app5}) in the limit $H \gg \Lambda_{\overline{MS}}$, while
the result for the opposite limit $H \ll \Lambda_{\overline{MS}}$
is in (4.97) of Ref.~\onlinecite{fat}:
\begin{equation}
{\cal F}_Q  = - \frac{H^2}{4 \pi c} \left[1 + \frac{1}{2 \ln
(2 \sqrt{2 \pi} e^{-3/4} \Lambda_{\overline{MS}}
/H)} \right] + \ldots~~~~~;~ H \ll \Lambda_{\overline{MS}},~~T=0.
\label{app6}
\end{equation}
Comparing (\ref{app6}) with (\ref{app4}), we immediately obtain
(\ref{e5}).

%%%%%%%%%%%%%%%%%%%%%%%%%%%%%%%%%%%%%%%%%%%%%%%%%%%%%%%%%%%%%%%%%%%%%%

\section{Short Time Expansion: Details}
\label{details}

We start with the expression (\ref{short2}) and insert (\ref{short3})
into it and the expand to first order to find
\begin{equation}
{\cal Z}_c= \int {\cal D \psi}{\cal D \psi^*}{\cal D \phi}{\cal D
\phi^*}\,
\exp(-\int dx \, {\cal L}),
\label{app-ste1}
\end{equation}
where
\begin{eqnarray}
{\cal L}& = & \psi\psi^* +
\frac{1}{2\xi} \left( \phi\psi^* + \phi^*\psi \right)^2 \nonumber \\
& + &  \left| \frac{\partial \phi}{\partial x} \right|^2
+ \frac{1}{2\xi} \left( \phi\frac{\partial \phi}{\partial x}^* +
\phi^*\frac{\partial \phi}{\partial x} \right)^2 \nonumber \\
&+ & m^2 \phi\phi^* + \frac{m^2}{2\xi} \phi^2 {\phi^*}^2 \nonumber \\
& - & \frac{1}{2 a \xi}\phi \phi^*.
\label{app-ste2}
\end{eqnarray}
Here we have dropped some additive constants and integrated by parts in places.
As advertised before, the magnetic field in the $z$ direction
parametrised by $m^2$ adds a mass term to the action and makes the
$\langle \phi \phi^* \rangle$ propagator infrared finite.  The last
term in the action arises from the Jacobian of the delta functionals
$\delta({\bf n}\cdot{\bf L})\delta({\bf n}^2-1)$ in the measure.  This Jacobian is
infinite for a continuum system.  To regularize it, we can introduce a
discrete lattice in space.  In that case the parameter $a$ is the
lattice constant, or equivalently the volume of the Brillouin zone.
\begin{equation}
\frac{1}{a} \equiv \frac{1}{L} \sum_{k \in \mbox{{\small BZ}}} 1
\label{app-ste3}
\end{equation}
This can be finite only in a finite volume system, but we shall carry
it through and ultimately it will cancel all ultraviolet divergences
arising from unrestricted momentum sums over the $\langle \psi\psi^*
\rangle$ propagator.  In terms of $\psi$ and $\phi$, the $\langle
{\bf L}(x,t)\cdot{\bf L}(0,0)\rangle$ is written as
\begin{eqnarray}
\langle {\bf L}(x,t)\cdot{\bf L}(0,0)\rangle& = &
\langle\psi(x,t)\psi^*(0,0)\rangle  + \mbox{c.c.} \nonumber \\
& + & \frac{1}{\xi}
\langle \phi(x,t)\psi^*(x,t)\phi(0,0)\psi^*(0,0) \rangle
+\mbox{c.c.} \nonumber \\
& + & \frac{1}{\xi}
\langle \psi(x,t)\phi^*(x,t)\psi^*(0,0)\phi(0,0) \rangle
+\mbox{c.c.}
\label{app-ste4}
\end{eqnarray}

Using (\ref{short3}) in the dynamical equations (\ref{short1}), we
find the following equations of motion:
\begin{eqnarray}
\frac{\partial \psi }{\partial t} & = &
i \left[
\frac{\partial^2 \phi}{\partial x^2}
%-m ^2\phi
+\frac{1}{\xi}\frac{\partial}{\partial
x}\left(\phi^2\frac{\partial\phi^*}{\partial x}\right)
\right] \nonumber\\
\frac{\partial \phi }{\partial t} & = & -i \left[ \psi +
\frac{1}{\xi}\left(\phi^2\psi^*\right)\right] .
\label{app-ste5}
\end{eqnarray}
These equations differ from (\ref{short7}) in that
we have absorbed a factor of $c=\xi^{1/2}$ into the
definition of time so that time and distance have the same units.

To solve for $\psi(x,t)$, we make a Fourier transform in space of
(\ref{app-ste5}) and convert it to an initial value problem given by
\begin{equation}
\frac{\partial}{\partial t}
\left[
\begin{array}{l}
\psi(k,t)\\
\phi(k,t)
\end{array}
\right]=\left[
\begin{array}{l}
-ik^2\phi(k,t)\\
-i\psi(k,t)
\end{array}
\right] + \left[
\begin{array}{l}
A(k,t)\\
B(k,t)
\end{array}
\right].
\end{equation}
We have here written the nonlinearities as the inhomogeneous part of a
set of first order equations.   The solution to the initial value
problem is given by
\begin{equation}
\left[
\begin{array}{l}
\psi(k,t)\\
\phi(k,t)
\end{array}
\right]=
{\cal K} (t) {\cal K} ^{-1}(0)
\left[
\begin{array}{l}
\phi(k,0)\\
\psi(k,0)
\end{array}
\right] +
{\cal K} (t)\int_0^t d\tau\, {\cal K} ^{-1}(\tau)\left[
\begin{array}{l}
A(k,\tau)\\
B(k,\tau)
\end{array}
\right].
\end{equation}
The columns of the $2\times2$ matrix ${\cal K} (t)$ is made up of the
two linearly independent solution vectors of the homogeneous problem.
\begin{equation}
{\cal K}(t)=\left[
\begin{array}{ll}
k e^{-ikt} & -k e^{ikt}\\
e^{-ikt} &  e^{ikt}
\end{array}
\right]
\end{equation}

Armed with this, we can solve for the fields in terms of the initial
conditions and iterate the solution to go to higher orders by plugging
the solution back in $A$ and $B$.   Also calculations simplify very
much if we work out everything back in real space.  So at last we
write down the final iterative form from which the entire perturbation
series can be generated.
\begin{eqnarray}
\psi(x,t) & = & \frac{1}{2}\left[\psi(x+t,0)+\psi(x-t,0)\right]+
\frac{i}{2}\frac{\partial}{\partial x}
\left[\phi(x+t,0)-\phi(x-t,0)\right] \nonumber \\
& + &
\frac{1}{2}\int_0^t d\tau \int dx^{\prime}
\left[
\delta(x-{x^{\prime}}+t-\tau)+\delta(x-{x^{\prime}}-t+\tau)
\right]A({x^{\prime}},\tau) \nonumber \\
& + &
\frac{i}{2}\int_0^t d\tau \int dx^{\prime}
\left[\delta(x-{x^{\prime}}+t-\tau) - \delta(x-{x^{\prime}}-t+\tau)
\right]
\frac{\partial}{\partial {x^{\prime}}}B({x^{\prime}},\tau)
\end{eqnarray}
\begin{eqnarray}
\phi(x,t) & = & \frac{1}{2}\left[\phi(x+t,0)+\phi(x-t,0)\right]-
\frac{i}{2}\int d{x^{\prime}}
\left[\theta(x-{x^{\prime}}+t)-\theta(x-{x^{\prime}}-t)\right]
\psi({x^{\prime}},0) \nonumber \\
& + &
\frac{1}{2}\int_0^t d\tau \int dx^{\prime}
\left[
\delta(x-{x^{\prime}}+t-\tau)+\delta(x-{x^{\prime}}-t+\tau)
\right]B({x^{\prime}},\tau) \nonumber \\
& - &
\frac{i}{2}\int_0^t d\tau \int dx^{\prime}
\left[\theta(x-{x^{\prime}}+t-\tau) - \theta(x-{x^{\prime}}-t+\tau)
\right]
A({x^{\prime}},\tau)
\end{eqnarray}
To simplify notation let us
denote the zeroth order solutions for the fields as $\psi^{(0)}(x,t)$
and $\phi^{(0)}(x,t)$.
\begin{eqnarray}
\psi^{(0)}(x,t)=&& \frac{1}{2}\left[\psi(x+t,0)+\psi(x-t,0)\right]+
\frac{i}{2}\frac{\partial}{\partial x}
\left[\phi(x+t,0)-\phi(x-t,0)\right] \nonumber \\
\phi^{(0)}(x,t)=&&\frac{1}{2}\left[\phi(x+t,0)+\phi(x-t,0)\right] \nonumber \\
&&~~~~~~-
\frac{i}{2}\int d{x^{\prime}}
\left[\theta(x-{x^{\prime}}+t)-\theta(x-{x^{\prime}}-t)\right]\psi({x^{\prime}},0)
\end{eqnarray}
Now we can proceed to evaluate each of the correlation functions in
(\ref{app-ste4}).  As an example let us look at
$\langle\psi(x,t)\psi^*(0,0)\rangle$.  The correlations to zeroth
order are
\begin{eqnarray}
\langle \psi^{(0)}(x,t){\psi^{(0)}}^*(0,0)\rangle & = & \frac{1}{2}\left[\delta(x+t)+
\delta(x-t)\right]\nonumber\\
\langle {\phi^{(0)}}(x,t){\phi^{(0)}}^*(0,0)\rangle & = & \frac{1}{4m}\left[e^{-m|x+t|}+
e^{-m|x-t|}\right]\nonumber\\
\langle {\phi^{(0)}}(x,t){\psi^{(0)}}^*(0,0)\rangle & = & \frac{i}{2}\left[\theta(x+t)
-\theta(x-t)\right]
\end{eqnarray}

To compute the one loop correlation
$\langle\psi(x,t)\psi^*(0,0)\rangle$, we write down
\begin{eqnarray}
\langle\psi(x,t)\psi^*(0,0)\rangle & =&
\langle\psi^{(0)}(x,t){\psi^{(0)}}^*(0,0)\rangle \nonumber \\
& + & \frac{1}{2}\int_0^t d\tau \int dx^{\prime}
\left[\delta(x-{x^{\prime}}+t-\tau)+\delta(x-{x^{\prime}}-t+\tau)\right]\times \nonumber\\
&& \frac{1}{\xi}\frac{\partial}{\partial {x^{\prime}}}
\left\langle\phi^2({x^{\prime}},\tau)
\frac{\partial\phi^*({x^{\prime}},\tau)}{\partial {x^{\prime}}}{\psi^{(0)}}^*(0,0)\right\rangle
\nonumber\\
&+& \frac{i}{2}\int_0^t d\tau \int dx^{\prime}
\left[\delta(x-{x^{\prime}}+t-\tau) - \delta(x-{x^{\prime}}-t+\tau)
\right]\times \nonumber\\
&& \frac{\partial}{\partial {x^{\prime}}}\left\langle-
\frac{i}{\xi}\phi^2({x^{\prime}},\tau)\psi^*({x^{\prime}},\tau){\psi^{(0)}}^*(0,0)\right\rangle
\end{eqnarray}
On the second and third terms of the RHS, we can replace
$\phi (\psi)$ by ${\phi^{(0)}}(\psi^{(0)})$ since
it is already first order in $1/\xi$.  Now all the correlation
functions on the RHS are expressed in terms of correlation functions
at time $t=0$.  So we evaluate them using the partition function
(\ref{app-ste1}).  Other correlations can be evaluated in the same
manner.  So we shall only write down the final answers here.
\begin{eqnarray}
\langle\psi(x,t)\psi^*(0,0)\rangle & = &
\frac{1}{2}\left[\delta(x+t)+\delta(x-t)\right]\left(1-\frac{1}{2m\xi}\right)\\
\langle \phi(x,t)\psi^*(x,t)\phi(0,0)\psi^*(0,0) \rangle & = &
\frac{1}{4\xi}\left[\theta(x+t)-\theta(x-t)\right]\\
\langle \psi(x,t)\phi^*(x,t)\psi^*(0,0)\phi(0,0)\rangle & = &
\frac{1}{8m\xi}\left[\delta(x+t)+\delta(x-t)\right]\left(1+e^{-2mt}\right).
\end{eqnarray}

Adding them all and taking the limit $m \rightarrow 0$ leads to the
expression quoted in (\ref{short4}).  Note that all terms which
diverge as $m \rightarrow 0$, cancel each other only when we evaluate
the ${\rm O}(3)$ invariant correlation $\langle {\bf L}(x,t) \cdot
{\bf L}(0,0)\rangle$.

%%%%%%%%%%%%%%%%%%%%%%%%%%%%%%%%%%%%%%%%%%%%%%%%%%%%%%%%%%%%%%%%%%%%%%

\section{Evaluation of the one-particle distribution function}
Let us start with equation \ref{jep7}.   Following Jepsen's notation we write
\begin{eqnarray}
&& P(y,t)= \sum_{k\neq 0}\langle \delta(y-x_k-v_kt)A_{0k}(t)\rangle+
\langle \delta(y-v_0t)A_{00}(t)\rangle  = \nonumber \\
&&\frac{1}{N} \sum_u \frac{N-1}{L} e^{-iju} \int_0^{L} d x_k
\int_{-\infty}^{\infty} d v_k g(v_k) Q(u,x_k+v_k t, v_k)
\left[ \frac{1}{L}\int_0^L dx_h R[u,x_k+v_kt-x_h]\right]^{N-2}
\nonumber\\
&&~~~ + \frac{1}{N} e^{-iju} \int_{-\infty}^{\infty} d v_0 g(v_0)
\delta(y-v_0 t)
\left[ \frac{1}{L}\int_0^L dx_h R[u,x_0+v_0t-x_h]\right]^{N-1} ,
\label{app-jep1}
\end{eqnarray}
with the definitions that
\begin{eqnarray}
g(v_k) & = & \frac{1}{2}(\delta(v_k-c)+\delta(v_k+c)),
\nonumber\\
Q((u,x_k+v_k t, v_k) &= & \int_{-\infty}^{+\infty} d v_0 g(v_0)
\delta(y - x_k- v_k t) S[u,x_k+v_kt-v_0t]
\nonumber \\
&=& \frac{1}{2}\delta(y - x_k- v_k t)\left(S[u,y-ct]+S[u,y+ct] \right),
\nonumber\\
R[u,x_k+v_kt-x_h]& = & \int_{-\infty}^{+\infty} d v_h\, g(v_h) S[u,x_k+v_kt-x_h-v_ht].
\end{eqnarray}

The second term in the RHS of (\ref{app-jep1}) clearly represents the
probability that the zeroth particle stays in the zeroth trajectory at
time $t$, while the first term represents the probability that it has
been scattered to another trajectory.  The second term can be easily
simplified using the definitions given above and in (\ref{jep1}).  It
is
\begin{eqnarray}
&& (\delta(y+ct)+\delta(y-ct))P^{(1)}(t) = \nonumber \\
&&~~~~~~~~~~~~\delta(y-ct) \frac{1}{2N}\sum_u
\left[ \frac{1}{2}\left(1+e^{ipu}\left(1+\frac{2ct-pL}{L}
(e^{iu}-1)\right)\right) \right]^{N-1}
\nonumber \\
&&~~~~~~~~~~~ + \delta(y+ct) \frac{1}{2N}\sum_u
\left[ \frac{1}{2}\left(1+e^{-ipu}\left(1+\frac{2ct-pL}{L}
(e^{-iu}-1)\right)\right) \right]^{N-1}
\end{eqnarray}
Here $p$ is defined so that
\begin{equation}
0<2ct-pL<L.
\end{equation}
$P^{(1)}(t)$ can't be evaluated in a closed analytic form.  But we
have evaluated it numerically for fixed values of $N$ and $L$.
$P^{(1)}(t)$ has been plotted in Fig \ref{fig9} for $N=50, L=1, c=\pm
1$.  As mentioned before, the peaks at times ${\cal T}, 2{\cal T},
3{\cal T}, \ldots$ are due to the exact recurrence of some
configurations.  We plot this function again for $N=50 $ and $51, L=1,
c=\pm 1$ in Fig \ref{fig10}.  Note that the peaks at times ${\cal T},
2{\cal T} \ldots $ disappear.

The first term will be handled in a manner similar to above.  For
simplicity we shall evaluate only the autocorrelation part.  To do
this, we set $y=nL$ where $n$ is an integer.
\begin{equation}
P^{(2)}(t)=\sum_{k\neq 0} \langle \delta(x_k+v_kt-nL)A_{0k}(t)\rangle .
\end{equation}
Using previous definitions we find that
\begin{eqnarray}
&& P^{(2)}(t)  =
\frac{1}{4N} \sum_u \frac{N-1}{L}
[\theta(nL-ct)-\theta(nL-ct-L)+\theta(nL+ct)-\theta(nL+ct-L)]
\times
\nonumber \\
&&    (S[u,nL+ct]+S[u,nL-ct])
\left[
\int_0^L \frac{dx}{2L} \,(S[u,nL-x-ct]+S[u,nL-x+ct])
\right]^{N-2}.
\end{eqnarray}
Here it will be convenient to define an integer $p$ as before such
that
\begin{equation}
0<ct-pL<L.
\end{equation}
Carrying out the integrals above we get
\begin{eqnarray}
P^{(2)}(t) & =  &
\frac{1}{4N} \sum_u \frac{N-1}{L} \left(e^{-iu}+e^{2ipu} \right)
\frac{1}{2^{N-2}} \nonumber\\
& \times & \left[
\frac{L-(ct-pL)}{L}\left(1+e^{2ipu}\right)+\frac{ct-pL}{L}\left(e^{-iu}+e^{i(2p+1)u}\right)
\right]^{N-2} + \mbox{c.c.}
\end{eqnarray}
The above function $P^{(2)}(t)$ has been plotted in Fig \ref{fig11}.

\end{document}